\documentclass[a4paper,12pt]{article}
\usepackage{amsmath}
\usepackage{amssymb,epsfig}
\usepackage{stmaryrd,wasysym,upgreek,latexsym}
\usepackage{lscape}
\usepackage[all]{xy}
\usepackage[vflt]{floatflt}
\usepackage{afterpage}
\usepackage[english]{babel}
\usepackage[verbose=true,twoside,hmarginratio=3:4]{geometry}

\newcommand{\qed}{{\hfill $\Box$}}
\newtheorem{lemma}{Lemma}[section]
\newtheorem{theorem}{Theorem}[section]

\newtheorem{definition}{Definition}[section]
\newtheorem{proposition}{Proposition}[section]
\newtheorem{example}{Example}[section]

\def\ZZ{{\mathchoice {\mathsf{Z \hspace{-0.45em} Z}} {\mathsf{Z 
        \hspace{-0.45em} Z}} {\mathsf{Z \hspace{-0.32em} Z}} 
    {\mathsf{Z \hspace{-0.23em} Z}}}}

\newcommand{\noi}{\nonumber}
\newcommand{\Om}{{\Omega}}

\newcommand{\e}{\varepsilon}
\newcommand{\beqa}{\begin{eqnarray}}
\newcommand{\eeqa}{\end{eqnarray}}

\newcommand{\be}{\begin{equation}}
\newcommand{\ee}{\end{equation}}
\newcommand{\bea}{\begin{eqnarray}}
\newcommand{\eea}{\end{eqnarray}}

\newcommand{\prf}{\noindent{\bf Proof}\ }

\newcommand{\Br}{\bar}

\newcommand{\ta}  {\tau}
\newcommand{\al}  {\alpha}
\newcommand{\lp}  {\left(}
\newcommand{\rp}  {\right)}

\newcommand\cF{{\mathcal F}}
\newcommand\cG{{\mathcal G}}

\newcommand\cR{{\mathcal R}}

\newcommand\cT{{\mathcal T}}

\numberwithin{equation}{section}
\begin{document}

\title{Topological Graph Polynomials\\ and Quantum Field Theory\\
Part I: Heat Kernel Theories}

\maketitle

\centerline{T. Krajewski$^{1,2}$, V. Rivasseau$^1$, A. Tanasa$^{3,4}$, Zhituo Wang$^{1,3}$}
\begin{center}
{\small\it 1) Laboratoire de Physique Th\'eorique, CNRS UMR 8627,\\
Universit\'e Paris XI,  F-91405 Orsay Cedex, France\\
2) on leave, Centre de Physique Th\'eorique, CNRS UMR 6207\\
CNRS Luminy, Case 907, 13288 Marseille Cedex 9\\
3) Centre de Physique Th\'eorique, CNRS UMR 7644,\\
Ecole Polytechnique F-91128 Palaiseau Cedex, France\\
4) IFIN, PO Box MG-6, 077125 Magurele, Romania\\
E-mail: krajew@cpt.univ-mrs.fr, rivass@th.u-psud.fr\\ adrian.tanasa@ens-lyon.org, ztwang@ens.fr}
\end{center}

\begin{abstract}

We investigate the relationship between the universal topological 
polynomials for graphs in mathematics and the parametric representation of Feynman amplitudes in quantum field theory.
In this first paper we consider translation invariant theories
with the usual  heat-kernel-based propagator.
We show how the Symanzik polynomials of quantum field theory
are particular multivariate versions of the Tutte polynomial, and how the new polynomials
of noncommutative quantum field theory are particular versions of the Bollob\'as-Riordan polynomials. 

\end{abstract}

\vfill
\noindent
LPT ORSAY 08-87\\
CPT-PXX-2008

\section{Introduction}
\setcounter{equation}{0}

Quantum field theory lies at the root of modern physics. 
After the success of the standard model in describing particle physics, 
one of the most pressing open question is how to derive an extended version
of field theory which encompasses the quantization of gravity. There are 
several attempts for this, among which string theory, loop gravity and noncommutative
geometry are the best known. In each of these attempts one of the key
problem is to relax the constraints that formulate quantum field theory 
on a particular space-time geometry.

What is certainly more fundamental than geometry is topology and in particular discrete
structures on finite sets such as the species of combinatorists \cite{BLL}. The most prominent 
such species in field theory is the species of Feynman graphs. 
They were introduced by Feynman to label quantum field perturbation theory and to
automatize the computation of {\it connected} functions. Feynman graphs 
also became an essential tool in \emph{renormalization}, the structure 
at the heart of quantum field theory.

There are two general canonical operations on graphs 
namely the deletion or contraction of edges. Accordingly perhaps the most important
quantity to characterize a graph is its Tutte polynomial \cite{Tutte,Crapo}. 
This polynomial obeys a simple recursion rule under
these two basic operations. It exists in many different variations, for instance 
multivariate versions, with possible decorations at vertices.
These polynomials have many applications, in particular to statistical physics.
For recent reviews see \cite{Sokal1,EM1,EM2}.

In recent years the Tutte polynomial has been generalized to the 
category of ribbon graphs, where it goes under the name of the Bollob\'as-Riordan polynomial
\cite{BR1,BR2,EM2}. Around the same time physicists have increasingly turned their attention
to quantum field theory formulated on noncommutative spaces, in particular
flat vector spaces equipped with the Moyal-Weyl product \cite{DouNe}.
This type of quantum field theory is hereafter called NCQFT.
It happens that perturbation theory for such NCQFT's is no longer labeled
by ordinary graphs but by ribbon graphs, suggesting a possible connection to the
work of Bollob\'as-Riordan.

Quantum field perturbation theory can be expressed in several representations. 
The momentum representation is the most common in the text books. The direct space
representation is closer to physical intuition.
However it is the parametric representation which is the most elegant and compact one.
In this representation, after the integration of internal position and/or momentum
variables has been performed explicitly, the result is expressed in 
terms of the Symanzik polynomials. There is an extensive literature on these polynomials (see e.g. \cite{Naka,IZ} for 
classical reviews).  These polynomials only depend on the Schwinger parameters.
Space time no longer enters explicitly into that representation except through its
dimension which appears simply as a parameter. 

This observation is crucial for several key applications in QFT which rely on dimensional
interpolation. Dimensional regularization and renormalization
was a crucial tool in the proof by 't~Hooft and Veltmann that non-Abelian gauge theories are renormalizable \cite{HV}. The Wilson-Fisher $\epsilon$ expansion \cite{Wil} is our
best theoretical tool to understand three dimensional phase transitions.
Dimensional regularization is also used extensively in the works of Kreimer and Connes 
\cite{Kreim,CKreim1} which recast the recursive BPHZ forest formula of perturbative renormalization into a Hopf algebra structure and relate it to a new class of Riemann-Hilbert
problems \cite{CKreim2}. 

Following these works, renormalizability has further attracted
considerable interest in the recent years as a pure mathematical
structure. The renormalization group ambiguity reminds
mathematicians of the Galois group ambiguity for roots of algebraic
equations \cite{Cartier}. Hence the motivations to study quantum field theory
and renormalization come no longer solely from physics but
also at least partly from number theory. 

The fact that the parametric representation is relatively independent of the details
of space time makes it also particularly appealing as a prototype for
the tools we need in order to quantize gravity.
The point of view of loop gravity is essentially based 
on the diffeomorphism invariance of general relativity.
In the spin foam or group field theory formalism amplitudes are expressed as discrete sums 
associated to combinatoric structures which generalize
Feynman graphs. They are in fact generalizations of ribbon graphs.
To extend the parametric representation and eventually the theory
of renormalization to this context is a major challenge, in which some 
preliminary steps have been performed \cite{Markopoulou}.

In this paper we uncover the relationship
between universal polynomials of  the Tutte and Bollob\'as-Riordan type
and the parametric representation in quantum field theory.  The Symanzik polynomials
that appear in ordinary commutative QFT are particular multivariate versions of Tutte polynomials.
The relation between Bollob\'as-Riordan polynomials and the non commutative analogs 
of the Symanzik polynomials
uncovered in \cite{MinSei,GurauRiv,RivTan} is new.
This establishes a relation between NCQFT, combinatorics and algebraic topology.
Recently the relation between renormalization and topological polynomials
was explored in \cite{KrajMart}, and in \cite{AluMarc}. We intend also to explore in the future
the relation between Feynman amplitudes and knot polynomials.

The plan of this paper is as follows. In the next section we 
give a brief introduction to graph theory and to Tutte-like 
polynomials. In the third section we derive the
parametric representation of Feynman amplitudes 
of QFT and give a new method to compute the 
corresponding Symanzik polynomials.
The deletion/contraction property (\ref{delcontrsym1}) of these polynomials
is certainly not entirely new \cite{Kreimer,brown}.
But our method which starts from the phase-space representation 
of Feynman amplitudes is inspired
by earlier work on NCQFT \cite{GurauRiv,RivTan} and introduces two main technical improvements. 
One is the use of Grassmann variables
to exploit the quasi-Pfaffian structure of Feynman amplitudes. This quasi-Pfaffian structure
was discovered in \cite{GurauRiv} in the context of NCQFT but 
to our knowledge was never applied to the simpler
case of ordinary QFT. The second improvement is that 
we do not factor out as usual the delta functions expressing 
global momentum conservation, because this requires a noncanonical choice
of a root for every connected graph. Instead we introduce an infrared regularization
in the form of a small harmonic potential at each vertex which  leads to more
elegant and canonical formulas. The corresponding generalized Symanzik polynomials obey
a transparent deletion/contraction relation which allows
to identify them with particular multivariate Tutte polynomials. These
polynomials are close but not identical to the polynomials of \cite{Sokal1};
we show how they both derive from a more general "categorified" polynomial.
The usual Symanzik polynomials are simply recovered as the leading terms
when the small harmonic potentials tend to zero.

For completeness we also include a more standard way to compute the Symanzik polynomials
through $x$ space representation and the tree matrix theorem.

In the fourth section we introduce ribbon graphs and Bollob\'as-Riordan polynomials.
In the fifth and last section we define the first and second Symanzik polynomials of NCQFT 
and relate them to the Bollob\'as-Riordan polynomials, using again the Pfaffian variables. 
Formulas for such polynomials were first sketched in \cite{MinSei}, but without proofs, 
nor relation to  the Bollob\'as-Riordan polynomials.

In a companion paper we shall discuss generalizations of the Tutte and Bollob\'as-Riordan
polynomials that occur for non-translation invariant theories with propagators
based on the Mehler rather than the heat kernel. These theories appeared as 
the first examples of renormalizable NCQFT's \cite{GrWu1,GrWu2,RVW,GMRV,Riv1}
and they are the most promising candidates for a fully non-perturbative
construction of a field theory in four dimensions \cite{GrWubeta,DR1,DGMR,Riv2}. In this case
the harmonic potentials on the vertices are no longer needed as the Mehler
kernel already contains an harmonic potential for the propagators of the graphs.

\section{Tutte Polynomial}
\setcounter{equation}{0}
\subsection{Graph Theory, Notations}
\label{graphsub}

A graph $G$ is defined as a set of vertices $V$ and of edges $E$ together with 
an incidence relation between them. The number of vertices and edges in a graph
will be noted also $V$ and $E$ for simplicity, since our context always prevents any confusion. 
Graph theorists and field theorists usually have different words for the same objects so
a little dictionary may be in order. We shall mostly use in this review 
the graph theorists language. In subsection \ref{decorsub} 
we introduce also some enlarged notion of graphs, with decorations called \emph{flags}
which are attached to the vertices of the graph to treat the external variables of 
physicists, plus other decorations also attached to vertices called (harmonic) weights to regularize infrared divergences.  Generalizations to ribbon graphs will be described in section \ref{briordan}.

\begin{figure}
\begin{center}
\includegraphics[scale=1,angle=-90]{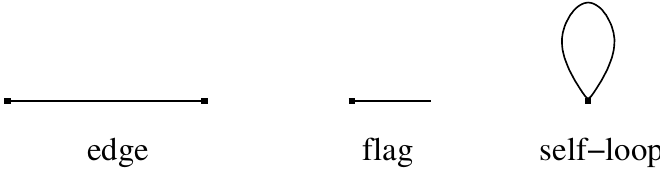}
\caption{Basic building blocks of a graph}
\end{center}
\label{fig:bas}
\end{figure}

Edges in physics are called lines (or propagators).
Edges which start and end at the same vertex are definitely allowed,
and called (self)-loops in graph theory and tadpoles in physics. 
A proper graph, i.e. a graph $G$ without such self-loops, together with
an arrow orienting each edge, 
can be fully characterized through its incidence
matrix ${\epsilon_{v e}}$.  It is the rectangular $E$ by $V$ matrix 
with indices running over vertices and
edges respectively, such that 

\begin{itemize}
\item
${\epsilon_{ve}}$ is +1 
if $e$  starts at $v$, 

\item
${\epsilon_{ve}}$ is -1 if $e$ ends at $v$,

\item
${\epsilon_{ve}}$ is  0 otherwise.

\end{itemize}

It is also useful to introduce the absolute value $\eta_{ve} = \vert \epsilon_{ve} \vert$ 
These quantities can be then generalized to graphs with self-loops by defining
$\epsilon_{ev} =0$ for any self-loop $e$ and vertex $v$ {\it but} $\eta_{ev} =2$
for a self-loop attached at vertex $v$ and $\eta_{ev} =0$ otherwise.
The number of half-edges at a vertex $v$ is called the degree of $v$
in graph theory, noted $d(v)$. Physicists usually call it the coordination number at $v$. 
A self-loop counts for 2 in the degree of its vertex, so that
$d(v) = \sum_{e} \eta_{ev}$.

An edge whose removal increases (by one) the number of connected parts 
of the graph is called a bridge in graph theory and a one-particle-reducible line in physics.

A forest is an acyclic graph and a tree is a connected forest.
A cycle in graph theory is a connected subset of $n$ edges and $n$ vertices 
which cannot be disconnected by removing any edge. It is called a loop in
field theory.

Physicists understood earlier than graph theorists that half-edges (also called flags
in graph theory \cite{Kauf}) are more fundamental than edges. This is because they correspond to integrated fields through the rule of Gau\ss ian integration, which physicists call Wick's theorem. 
Feynman graphs form a category of graphs with external flags decorating the vertices.
They occur with particular weights, in physics called amplitudes. These weights depend on the 
detail of the theory, for instance the space-time dimension. A quantum field theory can be viewed the 
generating functional for the species of such weighted Feynman graphs. In this paper we shall reserve the convenient word flag exclusively for the "external fields" decorations
and always use the word half-edge for the "internal half-edges".

An edge which is neither a bridge nor a self-loop is called regular. We shall call  \emph{semi-regular}
an edge which is not a self-loop, hence which joins two distinct vertices.

\begin{figure}
\begin{center}
\includegraphics[scale=0.9,angle=-90]{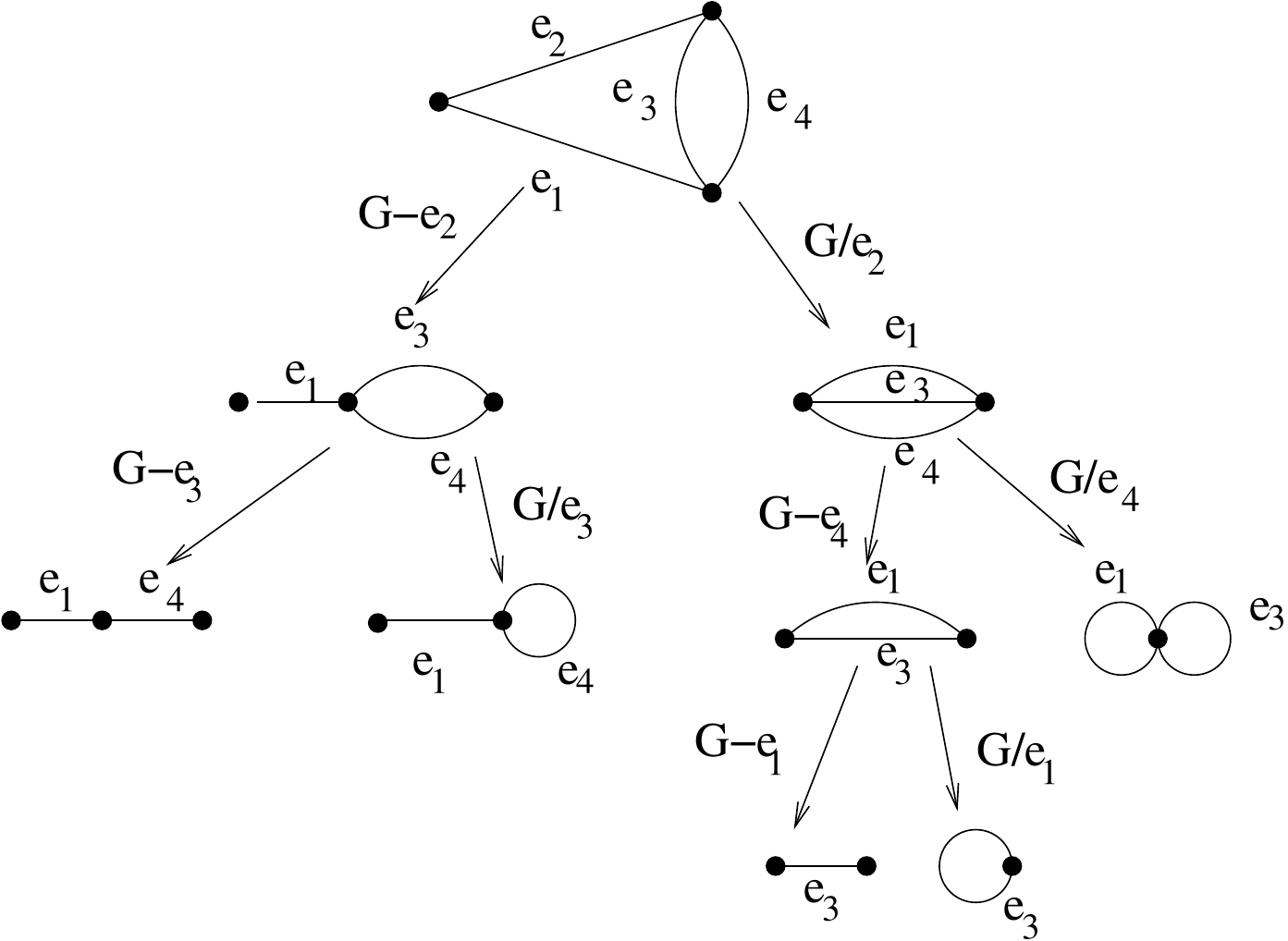}
\caption{The contraction-deletion of a graph}\label{fig:cond}
\end{center}
\end{figure}

There are two natural operations associated to an edge $e$
of a graph $G$, pictured in Figure \ref{fig:cond}:

\begin{itemize}
\item the deletion, which leads to a graph noted $G-e$,

\item the contraction, which leads to a graph noted $G/e$. If $e$ is not a self-loop, it identifies the two vertices $v_1$and $v_2$ at the ends of $e$ 
into a new vertex $v_{12}$, attributing all the 
flags (half-edges) attached to $v_1$ and $v_2$ to $v_{12}$, and then it removes $e$.
If $e$ is a self-loop, $G/e$ is by definition the same as $G-e$.

\end{itemize}

A subgraph $G'$ of $G$ is a
subset of edges of $G$, together with the attached vertices.
A \emph{spanning forest} of $G$ is an acyclic subgraph of $G$ 
that contains all the vertices of $G$. If $G$
is connected a spanning forest is in fact a 
tree of $G$ and any such \emph{spanning tree} has $\vert V\vert -1$ vertices.

As explained in the introduction
a topological graph polynomial 
is an algebraic or combinatoric object associated with a graph that is usually invariant under at least graph homeomorphism. It encodes information of the graph and so enables combinatoric and algebraic method to deal with graphs.

The Tutte polynomial \cite{Tutte} is one of the most general polynomial to
characterize a graph. It is defined under a simple rule through the
deletion and contraction of edges.  It can be generalized to the
larger theory of matroids \cite{Welsh}.

The original Tutte polynomial which is a function of two variables
can be generalized in various ways to multi-variable polynomials
which have many applications, in particular in statistical mechanics
where it evaluates the Potts model on graphs 
\cite{Sokal1,EM1,EM2}. These applications shall not be reviewed here.

We present first the two main equivalent definitions of the
Tutte polynomial. One direct way is to specify its linear recursion
form under contraction of regular edges (which are neither loops
nor bridges), together with an evaluation on terminal forms solely made
of bridges and self-loops. Another definition is as a rank-nullity generating 
function. By induction these definitions can be proved equivalent.

\subsection{Tutte Polynomial}

The definition through a recursion relation is a reduction rule on edges together
with an evaluation for the terminal forms.
The Tutte polynomial may be defined by such a linear recursion relation under deleting and
contracting regular edges. The terminal forms, i.e. the graphs without
regular edges are forests (i.e. graphs made of bridges) decorated with an additional arbitrary number of 
self-loops at any vertex. The Tutte polynomial evaluated on these terminal forms simply counts separately the number of bridges and loops:

\begin{definition}[Deletion-Contraction]\label{defdelcontr1}
If $G=(V,E)$ is a graph, and $e$ is a regular edge, then
\begin{equation}
T_G (x,y)=T_{G/e} (x,y)+T_{G-e} (x,y).
\end{equation}
For a terminal form $G$ with $m$ bridges and $n$ self-loops the polynomial is defined by
\begin{equation}
T_G(x,y)=x^m y^n .
\end{equation}
\end{definition}

It is not obvious that Definition \ref{defdelcontr1} is a definition at all since the result
might depend on the ordering in which different edges are suppressed through deletion/contraction, 
leading to a terminal form. The best proof that $T_G$ is unique and well-defined is in fact through
a second definition of the Tutte polynomial as a global sum over subgraphs. It gives a concrete
solution to the linear deletion/contraction recursion which is clearly independent
on the order in which edges are suppressed: 

\begin{definition}[Sum overs subsets]
 If $G=(V,E)$ is a graph, then the Tutte polynomial of
$G$, $T_G(x,y)$ has the following expansion:
\be
T_G (x,y)=\sum_{A\subset E}      (x-1)^{r(E)-r(A)} (y-1)^{n(A)},
\ee
where $r(A) =  \vert V \vert  - k(A) $ is the rank of the subgraph $A$ and 
$n(A) =\vert A \vert + k(A) - \vert V \vert  $ is its nullity or cyclomatic number.
In physicists language $n(A)$ is the number of independent loops in $A$.
\end{definition}

Remark that $r(A)$ is the number of edges in any spanning forest of $A$,
and $n(A)$ is the number of remaining edges in $A$ when a spanning forest
is suppressed, so it is the number of \emph{independent cycles} in $A$.

\begin{theorem}
These two definitions are equivalent.
\end{theorem}

One can show that the polynomial defined by the sum over subsets 
obeys the deletion-contraction recursion. One can also evaluate it directly
and show that it coincides with the first definition on the terminal forms
with only loops and bridges. 

There is a third definition of the Tutte polynomial through spanning trees 
(see eg \cite{EM1}). This third definition involves ordering the edges of the graph.
We think it may be also relevant in the context of field theory, in particular in relation
with the ordered trees or forests formulas of constructive theory \cite{BK,AR,GMR}, 
but this point of view will not be developed here.

\subsection{Multivariate Tutte polynomials}

Multivariate Tutte polynomials can also be defined through linear recursion or global formulas.

The ordinary multivariate Tutte polynomial 
$Z_G(q,\{\beta\})$ has a different variable $\beta_e$ for each edge $e$,
plus another variable $q$ to count vertices. We also write it 
most of the time as $Z_G(q,\beta)$ for simplicity.
It is defined through a completely general  linear deletion-contraction relation:
\begin{definition}[Deletion-Contraction]
For any edge $e$ (not necessarily regular)
\be\label{multivartut}
Z_G( q,\{ \beta \} )= \beta_e Z_{G/e} (q, \{\beta- \{\beta_e\} \} ) + Z_{G-e} (q, \{ \beta- \{ \beta_e  \} \} ) .
\ee
This relation together with the evaluation on terminal forms completely defines $Z_G(q,\beta)$,
since the result is again independent of the order of suppression of edges.
The terminal forms are graphs without edges,
and with $v$ vertices; for such graphs $Z_G(q,\beta)= q^v$.
\end{definition}

We can also define $Z_G(q,\beta)$ as a sum over subsets of edges:
\begin{definition}[Sum over subsets]
\begin{equation}
Z_G(q,\beta)=\sum_{A\subset  E}q^{k(A)}\prod_{e\in A}\beta_e ,
\end{equation}
where $k(A)$ is the number of connected components in the subgraph $(V,A)$.
\end{definition}
One can prove as for the two variables Tutte polynomial that 
this definition is equivalent to the first.
In \cite{Sokal1} this multivariate polynomial is discussed in detail.

\medskip
To understand the relation between this multivariate and the ordinary Tutte polynomial with two variables
we multiply $Z_G$ by $q^{- V } $, we set $ \beta_e = y-1$ and $q = (x-1)(y-1)$ and get
\bea
\big[ q^{- V }   Z_G(q,\beta) \big] \vert_{ \beta_e = y-1, q = (x-1)(y-1)} &=&
(x-1)^{k(E)  -|V|}  T_G(x,y).
\eea

We consider also
\be
q^{- k(G) }Z_G (q,\beta).
\ee 
Taking the limit $q \to 0$
that is retaining only the constant term in $q$
we obtain a sum over maximally spanning subgraphs $A$,
that is subgraphs with $k(A)=k(G)$:
\be S_{G} (\beta)=\sum_{A
\mathrm{ \; \; maximally  \; \; spanning  \; \; }  E  } \quad
\prod_{e\in A} \beta_e .
\ee

If we now retain only the lowest degree of homogeneity
in $\beta$ we obtain a sum over maximally spanning graphs
with lowest number of edges, ie maximally spanning acyclic graphs or 
\emph{spanning forests} of $G$.
\be F_{G} (\beta)=\sum_{\cF
\mathrm{ \; \; maximally  \; \; spanning  \; \;  forest    \; \;   of  \; \; }  G  } \quad
\prod_{e\in \cF} \beta_e .
\ee

Finally if we divide
$F_{G} (\beta)$ by $\prod_{e \in E} \beta_e$
and change variables to $\alpha_e = \beta_e^{-1}$ 
we obtain the ``(Kirchoff-Tutte)-Symanzik" polynomial. This polynomial is usually
defined for connected graphs, in which case the sum 
runs over spanning trees $\cT$
of $G$.
\begin{equation} \label{kts}
U_G(\alpha)=
\sum_{ \cT \; \;  \mathrm{spanning  \; \;  tree   \; \;   of  \; \; }  G  }\quad
\prod_{e\not\in \cT} \alpha_e .
\end{equation}

This polynomial satisfies the deletion contraction-recursion
\begin{equation}\label{delcontrsym1}
U_G(\alpha)=U_{G/e}(\alpha)+\alpha_e U_{G-e}(\alpha)
\end{equation}
for any regular edge $e$, together with the terminal form evaluation
\be \label{delcontrsym2}
U_G (\alpha) = \prod_{e   \; \;  \mathrm{self-loop}  \; \;  }   \   \ \ \alpha_e ,
\ee 
for any $G$
solely made of self-loops and bridges. 
The deletion-contraction (\ref{delcontrsym1}) can be extended
to general edges if we define $U$ for disconnected graphs as 
the product over the connected components of the corresponding 
$U$'s and put the contraction of any self-loop to 0.

The polynomial $U$ appears 
in a key computation of QFT, namely that of
the parametric representation of the Feynman amplitude associated to the graph $G$. 
We give a proof of this fact based on a new Pfaffian representation
of Feynman amplitudes together with harmonic weights at vertices so as 
to make the deletion/contraction rule (\ref{delcontrsym1})-(\ref{delcontrsym2}) 
particularly transparent.

But to define the second (Kirchoff-Tutte)-Symanzik polynomial as well as to make the computation
of the first Symanzik polynomial more canonical,
we need first to enlarge slightly our category of graphs
to include some decorations at the vertices.

\subsection{Decorated graphs}
\label{decorsub}

Decorations are essential  in physics to represent
the concept of {\it external variables}, which are ultimately those connected to 
actual experiments and observations. 

Graphs with integers attached to each vertex and their corresponding
multivariate polynomials $W_G ( \alpha_e, N_v)$ have been considered
in \cite{Wpolynomial}. But to represent external variables 
we need to replace the integer $N_v$ by a set of $N_v$
disjoint objects\footnote{In mathematics such a replacement is called a
categorification of the integers $N_v$.}, 
hereafter called \emph{flags} (see subsection \ref{graphsub}).

Each flag is attached to a single vertex. A momentum variable $p_f$
in ${\mathbb R}^d$ is associated to each such flag. The incidence matrix 
can be extended to the flags, that is we define
$\epsilon_{fv}$ as $+1$ if the flag $f$ is associated to the vertex $v$
and 0 otherwise. The total momentum incident to a subset $S$ of the 
graph is then defined as $\sum_{f} \sum_{v \in S}  \epsilon_{fv} p_f$.
Remark that this momentum is defined 
for subgraphs $S$ which may contain 
connected components reduced to single vertices.
For translation invariant QFT's,
global momentum conservation means that 
the condition $p_G =0$ must be fulfilled.

Similarly we attach to each vertex a number $q_v >0$ called the (harmonic) weight of the vertex.
The total weight of a subgraph $S$ is $\sum_{v \in S}  q_v$.

The deletion/contraction relation is then extended to this category of graphs.
The deletion is easy but the contraction is a bit non trivial. For a semi-regular edge joining
vertices $v_1$ and $v_2$ it collapses the two vertices into a single one $v_{12}$,
attaching to $v_{12}$ all half-edges of $v_1$ and $v_2$. But it also attaches to
$v_{12}$ the union of all the flags attached to $v_1$ and $v_2$, so that
the total momentum incoming to $v_{12}$ is the sum of the momenta incoming
to $v_1$ and to $v_2$. Finally the new weight of $v_{12}$ is the sum
$q_{v_1}+ q_{v_2}$ of the weights of $v_1$ and $v_2$.

These decorated graphs are the natural objects on which
to define generalized Symanzik polynomials in field theory.

Remaining for the moment in the context of graph theory we
can define the second (Kirchoff-Tutte)-Symanzik polynomial 
for a \emph{connected} graph as
\begin{definition}
\be  \label{secondsysy} V_G(\alpha, p)= - \frac{1}{2}\sum_{v \ne v'} p_v \cdot p_{v'}
\sum_{\cT_2 \; \;  \mathrm{2-tree \; \;separating } \;\: v  \;\; {\rm and }\;\; v`} \quad  
\prod_{e\not\in \cT_2} \alpha_e 
\ee
where a two tree $\cT_2$ means a tree minus one edge, hence a forest with 
two disjoint connected components $G_1$ and $G_2$; the separation condition
means that $v$ and $v'$ must belong one to $G_1$ the other to $G_2$.
\end{definition}

For any pair of distinct vertices $v$ and $v'$ we can build a canonical graph $G(v,v')$ first by joining vertices $v$ and $v'$ 
in $G$ with a new edge and then \emph{contracting that edge}. This operation could be called
the contraction of the pair of vertices $v$ and $v'$. The following result goes back to
Kirchhoff \cite{kirchhoff}.

\begin{proposition}\label{propUV} The second Symanzik polynomial is a quadratic form in the total momenta
$p_v$ at each vertex, whose coefficients are the $U_{G(v,v')}$ polynomials:
\be  V_G(\alpha, p)= -  \frac{1}{2}\sum_{v \ne v'} p_v \cdot p_{v'}\,\;  U_{G(v,v')} .
\ee
\end{proposition}
\prf The graph $G(v,v')$ has $V-1$ vertices, hence its spanning trees have $V-2$ edges. They cannot make
cycles in $G$ because they would make cycles in $G(v,v')$. They are therefore two-trees in $G$,
which must separate $v$ and $v'$, otherwise they would make  a cycle in $G(v,v')$.
\qed

On the submanifold of flag variables satisfying the \emph{momentum conservation}
condition $p_G = \sum_f p_f = 0$ there is an alternate less symmetric definition of a
similar polynomial:
\begin{definition}
\be\label{secondsy} \bar V_G(\alpha, p)=
\sum_{\cT_2 \; \;  \mathrm{2-tree }} \; \;   p_{G_1}^2  \; \; 
\prod_{e\not\in \cT_2} \alpha_e 
\ee
where $\cT_2$ is again a two-tree with two disjoint connected components $G_1$ and $G_2$.
\end{definition}
Indeed this is an unambiguous definition. On the submanifold $p_G =0$ we have
 $p_{G_1} = - p_{G_2}$, hence equation (\ref{secondsy})
does not depend of the choice of $G_1$ rather than $G_2$.

\begin{proposition}
On the manifold of flag variables satisfying the momentum conservation
condition $p_G = \sum_f p_f = 0$ one has $V_G(\alpha, p) = \bar V_G(\alpha, p)$.
\end{proposition}
\prf
We simply commute the sums over $v,v'$ and $\cT_2$
in (\ref{secondsysy}). For a given $\cT_2$ the condition that $v$ and $v'$ are separated allows
to separate the $p_v$ with $v \in G_1$ from the $p_{v'}$ with $v' \in G_2$; one gets therefore
$-  \frac{1}{2} \; 2 p_{G_1} \cdot p_{G_2} $ which is nothing but $p_{G_1}^2$
or $p_{G_2}^2$ on the manifold $p_G =0$.
\qed

We shall give in subsection \ref{symanpoly} a definition of
generalized first and second Symanzik polynomials for any graph, connected or not
from which $U_G$, $V_G$ or $\bar V_G$ can be easily derived in certain limits. 
Before actually performing these computations we include a brief interlude 
on Grassmann representation of determinants and Pfaffians. The reader familiar with this topic
can jump directly to the next section.

\subsection{Grassmann representations of determinants and Pfaffians}

Independent Grassmann variables $\chi_1, ..., \chi_n$ satisfy
complete anticommutation relations
\be \chi_i \chi_j = - \chi_j \chi_i \quad \forall  i,  j
\ee  so that any function of these variables is a polynomial
with highest degree one in each variable. 
The rules of Grassmann integrations
are then simply
\be  \int d\chi = 0,  \; \;    \;  \;  \int \chi d\chi = 1 .
\ee

The determinant of any $n$ by $n$ matrix can be then expressed as
a Grassmann Gau\ss ian integral over $2n$ independent
Grassmann variables which it is convenient to name as 
$\bar \psi_1, \ldots , \bar \psi_n$, $\psi_1, \ldots ,  \psi_n$,
although the bars have nothing yet at this stage to do with complex conjugation.
The formula is
\be \det M = \int \prod d\bar \psi_i d\psi_i   e^{-\sum_{ij} \bar\psi_i M_{ij} \psi_j   } .
\ee

The Pfaffian $\mathrm{Pf} (A)$ of an \emph{antisymmetric} 
matrix $A$ is defined by
\begin{equation}
\det A=[\mathrm{Pf} (A)]^2 .
\end{equation}

\begin{proposition}
We can express the Pfaffian as:
\begin{eqnarray}
\mathrm{Pf} (A) =\int d\chi_1...d\chi_n
e^{-\sum_{i<j}\chi_i A _{ij}\chi_j}
= \int d\chi_1...d\chi_n e^{-\frac{1}{2}\sum_{i,j}\chi_i A _{ij}\chi_j} .
\label{pfaff}
\end{eqnarray}
\end{proposition}

\prf Indeed we write
\begin{equation}
\det A= \int \prod_i d\bar \psi_i d\psi_i   e^{-\sum_{ij} \bar\psi_i A_{ij} \psi_j   }  .
\end{equation}
Performing the change of variables (which a posteriori justifies the complex notation)
\begin{eqnarray} \label{changepfaff}
\bar\psi_i = \frac{1}{ \sqrt{2}}(\chi_i - i\omega_i), \quad 
\psi_i = \frac{1}{\sqrt{ 2}}(\chi_i + i\omega_i),
\end{eqnarray}
whose Jacobian is $i^{-n}$, the new 
variables $\chi$ and $\omega$ are again independent Grassmann variables.
Now a short computation using $A_{ij}=-A_{ji}$ gives
\bea
\det A&=&  i^{-n}  \int \prod_i d\chi_i d\omega_i   e^{-\sum_{i<j} \chi_i A_{ij} \chi_j 
- \sum_{i<j} \omega_i A_{ij} \omega_j  } \nonumber \\
&=&  \int \prod_i d\chi_i  e^{-\sum_{i<j} \chi_i A_{ij} \chi_j  }\prod_i  d\omega_i   e^{ -\sum_{i<j} \omega_i A_{ij} \omega_j  },
\label{pfaffi}\eea
where we used that $n=2p$ has to be even and that a factor $(-1)^p$ is generated
when changing $ \prod_i d\chi_i d\omega_i $ into $ \prod_i d\chi_i \prod_i d\omega_i $.
Equation (\ref{pfaffi}) shows why $\det A$ is a perfect square and proves (\ref{pfaff}). \qed

\begin{lemma}\label{quasipfaff}
The determinant of a matrix $D+A$ where $D$ is
diagonal and $A$ antisymmetric has a "quasi-Pfaffian" representation
\be  \det (D+A) = \int \prod_i d\chi_i d \omega_i e^{-\sum_i  \chi_i D_{ii} \omega_i - \sum_{i <j}
\chi_i A_{ij} \chi_j + \sum_{i <j}  \omega_i A_{ij} \omega_j } .
\ee
\end{lemma}
\prf  The proof consists in performing the change of variables 
(\ref{changepfaff}) and canceling carefully the $i$ factors. \qed

\subsubsection{Tree-Matrix Theorem}

Let $A$ be an $n\times n$ matrix such that 
\be \label{sumnulle}
\sum_{i=1}^n A_{ij} = 0 \ \ \forall j \ .
\ee Obviously $\det A=0$. The interesting quantities are eg the 
diagonal minors $\det A^{ii}$ obtained by 
deleting the $i$-th row and  the $i$-th column in $A$. The ``Kirchoff-Maxwell"
matrix tree theorem expresses these minors as sums over trees:

\begin{theorem}[Tree-matrix theorem]
\be\label{treemat}
\det A^{ii}= \sum_{T\ {\rm spanning\  tree\ of} A} \prod_{e \in T} (-A_e ) ,
\ee
where the sum is over spanning trees on $\{1, ... n\}$ oriented away from root $i$.
\end{theorem}

\prf We give here a sketch of the Grassmann proof given in \cite{Abd}.
We can assume without loss of generality that $i=1$. For any matrix A we have:
\begin{equation}
\det A^{11}= \int    \big[ \prod_{i=1}^n d\bar\psi_i d\psi_i \big] \psi_1 \bar\psi_1  e^{-\sum_{i,j}\bar\psi_i
A_{ij}\psi_j} .
\end{equation}
The trick is to use (\ref{sumnulle}) to write
\be
{\Br \psi}A\psi=\sum_{i,j=1}^n
({\Br\psi}_i-{\Br\psi}_j)A_{ij}\psi_j ,
\ee
hence
\bea
\det A^{11} &=&
\int {\rm d}{\Br\psi} {\rm d}\psi
\ (\psi_1 {\Br \psi}_1)
\exp \lp -\sum_{i,j=1}^n A_{ij}({\Br\psi}_i-{\Br\psi}_j)\psi_j
\rp
\nonumber\\
&=& \int {\rm d}{\Br\psi} {\rm d}\psi
\ (\psi_1 {\Br \psi}_1)
\left[ \prod_{i,j=1}^n \lp 1-A_{ij}({\Br\psi}_i-{\Br\psi}_j)\psi_j \rp
\right]
\eea
by the Grassmann rules. We now expand to get
\be
\det A^{11} =
\sum_{\cG}
\lp
\prod_{\ell=(i,j)\in\cG}(-A_{ij})
\rp
\Om_{\cG}
\ee
where $\cG$ is {\em any} subset of $[n]\times[n]$, and we used the notation
\be
\Om_{\cG}\equiv
\int {\rm d}{\Br\psi} {\rm d}\psi
\ (\psi_1 {\Br \psi}_1)
\lp
\prod_{(i,j)\in\cG}
\left[ ({\Br\psi}_i-{\Br\psi}_j)\psi_j \right]
\rp .
\ee

Then the theorem follows from the following
\begin{lemma}
$\Om_{\cG}=0$
unless the graph $\cG$
is a tree directed away from 1 in which case
$\Om_{\cG}=1$.
\end{lemma}

\prf
Trivially, if $(i,i)$ belongs to $\cG$, then the integrand of
$\Om_{\cG}$ contains a factor ${\Br\psi}_i-{\Br\psi}_i=0$ and
therefore $\Om_{\cG}$ vanishes. 

But the crucial observation is that if 
there is a loop in $\cG$ then again $\Om_{\cG}=0$.
This is because then the integrand of $\Om_{\cF,\cR}$ contains the factor
\be
{\Br\psi}_{\ta(k)}-{\Br\psi}_{\ta(1)}=
({\Br\psi}_{\ta(k)}-{\Br\psi}_{\ta(k-1)})+\cdots+
({\Br\psi}_{\ta(2)}-{\Br\psi}_{\ta(1)}) .
\ee
Inserting this telescoping expansion of the factor
${\Br\psi}_{\ta(k)}-{\Br\psi}_{\ta(1)}$ into the integrand of 
$\Om_{\cF,\cR}$, the latter breaks into a sum of $(k-1)$ products.
For each of these products, there exists an $\al\in\ZZ/k\ZZ$
such that the factor $({\Br\psi}_{\ta(\al)}-{\Br\psi}_{\ta(\al-1)})$
appears {\em twice} : once with the $+$ sign from the telescopic
expansion of $({\Br\psi}_{\ta(k)}-{\Br\psi}_{\ta(1)})$, and once more
with a $+$ (resp. $-$) sign if $(\ta(\al),\ta(\al-1))$
(resp. $(\ta(\al-1),\ta(\al))$) belongs to $\cF$.
Again, the Grassmann rules entail that $\Om_{\cG}=0$. \qed

To complete the proof of (\ref{treemat}) every connected component of $\cG$ must contain 
1, otherwise there is no way to saturate the $d\psi_1$ integration. 

This means that $\cG$ has to be a directed tree on $\{1,... n\}$.
It remains only to see that $\cG$ has to be directed away from 1,
which is not too difficult.
\qed

The interlude is over and we now turn to perturbative QFT and to the
parametric representation of Feynman amplitudes.

\section{Parametric Representation of Feynman Amplitudes}
\setcounter{equation}{0}

In this section we will give a brief introduction to the
parametric representation of ordinary QFT on a commutative
vector space ${\mathbb R}^d$. We may take the example of $\phi^4$ bosonic theory but
the formalism is completely general.

\subsection{Green and Schwinger functions in QFT}

In particle physics the most important quantity is the diffusion
matrix S whose elements or cross sections can be measured in
particle experiments. The S matrix can be expressed from the Green
functions through the reduction formulas. Hence they
contain all the relevant information for that QFT.

These Green functions are time ordered vacuum expectation values of the
fields $\phi$, which are operator-valued and act on the Fock space:
\begin{equation}
G_N(z_1,...,z_N)=\langle
\psi_0,T[\phi(z_1)...\phi(z_N)]\psi_0\rangle .
\end{equation}
Here $\psi_0$ is the vacuum state and the $T$-product orders
$\phi(z_1)...\phi(z_N)$ according to increasing times.

In the functional integral formalism the Green functions can be
written as:
\begin{equation}
G_N(z_1,...,z_N)=\frac{\int\prod_{j=1}^N\phi(z_j){e^{i\int\mathcal
{L}(\phi(x))dx}}D\phi}{\int{e^{i\int\mathcal {L}(\phi(x))dx}}D\phi} .
\end{equation}

Here $\mathcal {L}=\mathcal {L}_0+\mathcal {L}_{int}$ is the full
Lagrangian of the theory. The Green functions continued to Euclidean
points are called the Schwinger functions and are given by
the Euclidean Feynman-Kac formula:
\begin{equation}
S_N(z_1,...,z_N)=Z^{-1}\int\prod_{j=1}^N\phi(z_j)e^{-\int\mathcal
{L}(\phi(x))dx}D\phi ,
\end{equation}
\begin{equation}
Z=\int e^{-\int\mathcal {L}(\phi(x))dx}D\phi .
\end{equation}
For instance for the $\phi^4$ theory,  $\mathcal {L}_{int}=\frac{\lambda}{4!} {\phi (x)}^4$
and we have
\begin{equation}\label{phi4theory}
\mathcal {L}(\phi)=\frac{1}{2} \partial_\mu \phi(x) \partial^\mu
\phi(x) +\frac{1}{2}m{\phi(x)}^2 + \frac{\lambda}{4!} {\phi (x)}^4
\end{equation}

where
\begin{itemize}
\item $\lambda$ is the (bare) coupling constant, which characterizes the
strength of the interaction, the traditional factor 1/4! is inessential but slightly 
simplifies some computations.

\item $m$ is the (bare) mass,

\item $Z$ is the normalization factor,

\item $D\phi$ is an ill-defined "flat" product of Lebesgue measures $\prod_x d\phi(x)$ 
at each space time point.
\end{itemize}

The coefficient of the Laplacian is set to 1 in (\ref{phi4theory}) for simplicity. Although this coefficient
actually in four dimensions flows through renormalization, it is possible to exchange this flow 
for a rescaling of the field $\phi$.

To progress towards mathematical respectability and to prepare for perturbation theory,
we combine the $e^{-\int\mathcal {L_0}(\phi(x))dx}D\phi$ and the free normalization factor
$Z_0 = \int e^{-\int\mathcal {L_0}(\phi(x))dx}D\phi$ into a normalized Gau\ss ian measure $d\mu_C (\phi)$
which is well-defined on some subspace of the Schwartz space of distributions $S'(R^d)$ \cite{GJ}.
The covariance of this measure is the (free) translation invariant propagator 
$C(x,y) = \int \phi(x) \phi(y) d\mu_C (\phi)$, which by slight abuse of notation
we also write as $C(x-y)$ and whose Fourier transform is
\begin{equation}
C(p)=\frac{1}{(2\pi)^d}\frac{1}{p^2+m^2}.
\end{equation}

In this way the Schwinger functions are rewritten as
\begin{equation}
S_N(z_1,...,z_N)=Z^{-1}\int_{R^d}\prod_{j=1}^N \phi(z_j)  e^{ -\int_{R^d}   \mathcal{L}_{int}   (\phi) }   d\mu_C(\phi ),
\end{equation}
\begin{equation}
Z=\int e^{-\int_{R^d} \mathcal {L}_{int}(\phi(x))dx}  d \mu_C(\phi).
\end{equation}

However this expression is still formal for two reasons; for typical fields the interaction factor is
not integrable over $R^d$ so that $\int_{R^d}   \mathcal{L}_{int}   (\phi)$ is ill-defined (infrared or thermodynamic problem)
and in dimension more than 2 even when the interaction factor is restricted to a finite volume
it is still ill-defined because for typical distributions $\phi$, products such as $\phi^4(x)$
are also ill-defined. This is the famous ultraviolet problem which requires renormalization (see \cite{Riv3}), 
but this problem is not addressed here, as we discuss solely the structure of the integrands in Feynman 
parametric representations, not the convergence of the integrals. 
The reader worried by ill-defined integrals in the rest of this paper
for space-time dimension $d$ larger than $2$ should impose a
ultraviolet regulator. This means he should replace replace $C(p)$ 
by a better behaved $C_{\kappa}(p)$ such as
\begin{equation}
C_\kappa(p)=\frac{1}{(2\pi)^d}\frac{e^{-\kappa(p^2+m^2)}}{p^2+m^2}=\int_\kappa^\infty
e^{-\alpha (m^2+p^2)}d\alpha ,
\end{equation}
so that
\begin{equation}
C_\kappa(x,y)=\int_\kappa^\infty e^{-\alpha
m^2-(x-y)^2/{4\alpha  }}\frac{d\alpha}{\alpha ^{D/2}} .
\end{equation}

We now turn to perturbation theory in which the factor $e^{-\int_{R^d}   \mathcal{L}_{int}   (\phi)  }$
is expanded as a power series. This solves the thermodynamic problem, at the cost of introducing 
another problem, the divergence of that perturbation expansion. This divergence which in the good cases
can be tackled by constructive field theory \cite{GJ,Riv4,constr1,constr2} will not be treated in this paper.

\subsection{Perturbation theory, Feynman Graphs}

Wick theorem is nothing but the rule of pairing which computes the moments of
a Gau\ss ian measure. It allows to integrate monomials of fields
\begin{equation}
\int \phi(x_1)...\phi(x_n)d\mu_C(\phi)=\sum_G \prod_{e\in G}C(x_{i_e},x_{j_e})
\end{equation}
where the sum over $G$ is over all contraction schemes (i.e. pairings of the fields) and 
$C(x_{i_e},x_{j_e})$ is the propagator kernel joining the arguments of the two fields 
$\phi(x_{i_e})$ and $\phi(x_{j_e})$  paired
into the edge $e$ by the contraction scheme $G$.

It was Feynman's master stroke to represent each such contraction scheme by a particular \emph{graph}
in which edges represent pairs of contracted fields and vertices stand for  the interaction. 

In the case of a $\phi^4$ theory, remark that these interaction vertices have degree 4.
Indeed the Schwinger functions after perturbative expansion are
\begin{equation}
S_N(z_1...z_N)=\frac{1}{Z}\sum_{n=0}^\infty\frac{(-\lambda)^n}{4^n n!}
\int\big[\int    \prod_{v=1}^n  \phi^4(x_v)dx_v \big]  \phi(z_1)...\phi(z_N)d\mu(\phi).
\end{equation}
The pairings of Wick's theorem therefore occur between $n$ internal vertices each equipped with four fields and
$N$ external vertices or sources corresponding to the single fields  $\phi(z_1)$, ... , $\phi(z_N)$.

Schwinger function are therefore expressed as sums over Feynman graphs of 
associated quantities or weights called the Feynman amplitudes. In this position space 
representation the Feynman graphs
have both $n$ \emph{internal vertices} corresponding to the 
${\cal L}_{int}$ factors, plus $N$ external vertices of degree 1 corresponding to the fields 
$\phi(z_1), ... ,  \phi(z_N)$. In the case of the $\phi^4$ theory
each internal vertex has degree 4.

\begin{figure}
\begin{center}
\includegraphics[scale=0.9,angle=-90]{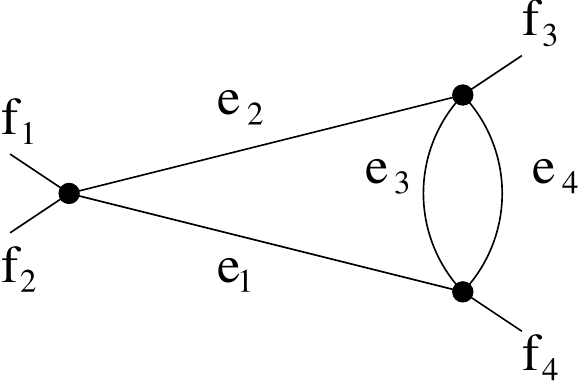}
\caption{ A $\phi^4$ graph}\label{fig:nor}
\end{center}
\end{figure}

The Feynman amplitudes are obtained by integrating over all positions of internal vertices
the product of the propagator kernels for all the edges of the graphs
\be \label{amplix}
A_G (z_1, ..., z_N) =  \int   \prod_{v} dx_v   \prod_{e \in G}  C(x_{i_e},x_{j_e}) ,
\ee
where the product $\prod_v$ runs over the \emph{internal} vertices $v$.

The quantities that are relevant to physical experiments are the
\emph{connected} Schwinger functions which can be written as:
\begin{equation}
\Gamma_N(z_1,...,z_N)=\sum_{ \phi^ 4 {\rm \ connected\  graphs\   } G {\rm \ with\ }
N(G)=N  }    \frac{(-\lambda)^{n(G)}}{S(G)} A(G)(z_1,...,z_N),
\end{equation}
where $S(G)$ is a combinatoric factor (symmetry factor).

The momentum space representation corresponds
to a Fourier transform to momenta variables called $p_1,... , p_N$: 
\be \Gamma_N(p_1,...,p_N) =\int dz_1...dz_N
e^{2i\sum p_f z_f} \Gamma_N(z_1,...,z_N),
\ee
where the factor 2 is convenient and we forget inessential normalization factors.
This is a distribution, proportional to a global momentum conservation
$\delta (\sum_{f=1}^N p_f)$. From now on we use an index $f$ to label external momenta 
to remember that they are associated to corresponding 
graph-theoretic \emph{flags}. Usually one factors out this
distribution together with the external propagators, to obtain the 
expansion in terms of truncated amputated graphs:

\bea
\Gamma^T_N(p_1,...,p_N)&=&
\sum_{\phi^4 {\rm \ truncated\ graphs\  } G {\rm \ with\ }
N(G)=N}\frac{(-\lambda)^{n(G)}}{S(G)}
\nonumber \\
&&\delta (\sum_{f=1}^N p_f)
\prod_{f=1}^N   \frac{1}{p_f^2 + m^2} A^T_G(p_1,...,p_N) .
\label{globaldelta}
\eea
In this sum we have to describe in more detail
the \emph{truncated graphs} $G$ with $N$ external flags.
Such truncated graphs are connected, but they may contain bridges and self-loops. They no longer have external vertices of degree 1. Instead, they still have $N$ external variables $p_f$, no longer 
associated to edges but to flags ($N$ in total), which decorate the 
former internal vertices. For instance for the $\phi^4$ theory 
the degree of a truncated graph $G$ is no longer 4 at each 
internal vertex. It is the total degree, that is the number of half-edges plus flags which 
remains 4 at every vertex.

\begin{figure}
\begin{center}
\includegraphics[scale=0.9,angle=-90]{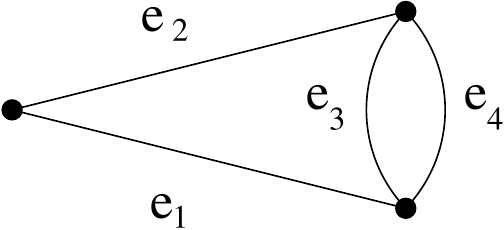}
\caption{ A truncated $\phi^4$ graph}\label{fig:trun}
\end{center}
\end{figure}
Ordinary Schwinger functions can be expressed as sums over partitions of the arguments of products of the corresponding
truncated functions. We now give the explicit form of the corresponding 
truncated amplitudes $A^T_G(p_1,...,p_N)$.

\subsection{Parametric representation}\label{paramQFT}

We shall first consider a fixed truncated oriented diagram $G$ 
and compute the corresponding contribution 
or amplitude $A^T_G$ as given by Feynman rules.

We denote again by $E$ and $V$ the
number of edges and vertices respectively, and by $N$ the number
of flags. Since $G$ is connected its incidence matrix has rank $V-1$.

Now consider a Feynman graph $G$ contributing to some truncated
Schwinger function $\Gamma^T(p_1,...,p_N)$. The usual way to take into 
account the global $\delta$ function
in (\ref{globaldelta}) is to restrict to configurations such that $\sum_f  p_f=0$.
Extraction of this global delta function in (\ref{globaldelta}) for the amplitude
of a particular graph can be done provided we do not integrate
the position of one of the vertices in (\ref{amplix}),  but rather fix it
at an arbitrary point, eg the origin. From now on we suppose
this vertex is $\bar v$ the one with last index. It provides a \emph{root}
in the graph $G$. However this standard procedure requires the non-canonical choice
of that root vertex, and the final result does not depend on that choice.

Another possibility is to modify the interaction $\lambda \phi^{4}(x)$
into $\lambda e^{-q x^2}   \phi^{4}(x)$, in which case there is no longer global momentum conservation.
One can compute modified amplitudes $B^T_G (p_1, ... p_N; q)$ 
without factoring out the global $\delta (\sum_{f=1}^N p_f)$ factor, so that
\bea
\Gamma^T_N(p_1,...,p_N; q)&=&
\sum_{\phi^4 {\rm \ truncated\ graphs\  } G {\rm \ with\ }
N(G)=N}\frac{(-\lambda)^{n(G)}}{S(G)}
\nonumber \\
&&
\prod_{f=1}^N   \frac{1}{p_f^2 + m^2} B^T_G(p_1,...,p_N; q) .
\label{globaldelta1}
\eea
The momentum conserving usual amplitudes are recovered when $q \to 0$:
\be\label{limBA} \lim_{q \to 0}   B^T_G(p_1,...,p_N; q)  = \delta (\sum_{f=1}^N p_f)  A^T_G  (p_1, ... ,p_N) .
\ee
This is the procedure we shall follow
in subsection \ref{symanpoly}, because it avoids the choice of a noncanonical root.
But for the moment let us complete the standard presentation of $A^T_G  (p_1, ... ,p_N)$.

The momentum representation of $A^T_G$, forgetting from now on
inessential factors of $2\pi$, is:
\begin{eqnarray}
A^T_G  (p_1, ... ,p_N)&=& \int\prod_{e=1}^E d^d k_e
\frac{1}{k_e^2+m^2} 
\prod_{v=1}^{V-1} \delta (\epsilon_{fv}  p_f+\epsilon_{ev}k_e) .
\label{ampli}
\end{eqnarray}
in which we use the convention that repeated indices are summed,
so that $\epsilon_{fv}  p_f+\epsilon_{ev}k_e$
stands for the total momentum $\sum_f \epsilon_{fv}  p_f+ \sum_e \epsilon_{ev}k_e$
incoming at vertex $v$.

To obtain the parametric representation we have first to
rewrite the propagators as :
\begin{equation} \frac{1}{k^2+m^2}=\int_0^{\infty} d\alpha
e^{-\alpha(k^2+m^2)} .  \label{prop1}
\end{equation}
We obtain the momentum parametric representation
\begin{equation}
A^T_G(p_1,...,p_N)=\int \prod_{e =1}^E    d \alpha_e d^d
k_e e^{- \alpha_e (k_e^2+m^2)} 
\prod_{v=1}^{V-1}  \delta (\epsilon_{fv}  p_f+\epsilon_{ev}k_e) .
\label{paramoment}
\end{equation}

Fourier transforming the $V-1$ Dirac distributions into oscillating integrals
we obtain, up to some inessential global factors the phase-space parametric representation
\begin{equation}
A^T_G(p_1,...,p_N) = \int  \prod_{e =1}^E \big[ d \alpha_e  e^{-\alpha_e m^2} d^d k_e  \big]
 \prod_{v=1}^{V-1} d^d x_v
e^{-\alpha_e k_e^2  + 2 i ( p_f  \epsilon_{fv} x_v   +  k_e\epsilon_{ev} x_v ) } ,
\label{paraphase}
\end{equation}
where again
$ k_e\epsilon_{ev} x_v$ means
 $\sum_{e=1}^E  \sum_{v=1}^{V-1}
k_e\epsilon_{ev} x_v$ etc, and the factor 2 is convenient. 

Finally integrating out the edge momenta whose dependence is Gau\ss ian
leads to the $x$ or direct space parametric representation:

\begin{equation}
A^T_G(p_1,...,p_N)=  \int \prod_{e =1}^E   d \alpha_e  
\frac{e^{-\alpha_e  m^2}}{\alpha_e^{d/2}} 
\prod_{v=1}^{V-1} d^d x_v    e^{2i p_f  \epsilon_{fv} x_v   -  x_v \cdot x_{v'} \epsilon_{ve} \epsilon_{v'e}  / \alpha_e  }  .
\label{paradirect}
\end{equation}
Remember this amplitude is only defined on the submanifold $p_G =0$, because
it is only there that the formula gives a result independent of the choice of the
root not integrated out in (\ref{paradirect})

The parametric representation consists in integrating out fully
the $x$ or $p$ variables in (\ref{paramoment}), (\ref{paraphase})
or (\ref{paradirect}).
One obtains the parametric representation, which is an integral
on $\alpha$ parameters only:
\begin{equation}
A^T_G(p_1,...,p_N) = \int  \prod_{e =1}^E  \big[ d \alpha_e e^{- \alpha_e m^2}\big]
\frac{e^{-V_G (p, \alpha)/  U_G (\alpha)}}{U_G (\alpha)^{d/2}} ,
\label{paramet}
\end{equation}
where $U_G$ and $V_G$ are called the first and second 
Symanzik's polynomials.  

\begin{theorem}\label{theosym}
The first Symanzik polynomial $U_G$ in (\ref{paramet})
is the multivariate Tutte polynomial (\ref{kts}).
On the submanifold $p_G =0$, the only one where it is unambiguously defined, 
the second polynomial $V_G$  of (\ref{paramet}) coincides with (\ref{secondsysy}) and (\ref{secondsy}).
\end{theorem}

We are going to give two proofs of this classic theorem of quantum field theory,
one relying directly on contraction-deletion and on the phase-space representation  (\ref{paraphase})
the other more standard and relying on the direct representation  (\ref{paradirect}) and on the tree-matrix theorem.

Indeed in order to compute the Symanzik's polynomials, let us remark first that the
momentum representation mostly used in textbooks
is not very convenient. To use (\ref{paramoment})
we should ``solve" the $\delta$ functions, that is rewrite
each edge momentum in terms of independent momenta for cycles. In physics
this is called a momentum routing. But such a momentum routing is linked to the
choice of a particular spanning tree of $G$. The momenta of the edges not in this tree are kept as independent variables and the tree edges momenta are recursively computed in terms of those by progressing from the leaves of the  tree towards the root which is the fixed vertex $v_n$. This is not a canonical prescription, as it depends on the choice of the tree.

The representations (\ref{paraphase})
or (\ref{paradirect}) are more convenient to integrate the space or momentum variables 
because the dependence in variables $x$ and $k$ is Gau\ss ian so that the result
is a determinant to a certain power times a Gau\ss ian in the external variables.
In fact  (\ref{paraphase}) is the best as we shall argue below.
However there is still a small noncanonical choice, the one of the root. This is why
we prefer to compute the regularized amplitudes
\begin{equation}
B^T_G(p_1,...,p_N;q ) = \int  \prod_{e =1}^E  \big[
d \alpha_e e^{-\alpha_e m^2}   d^d k_e  \big]  \prod_{v=1}^{V} d^d x_v
e^{-\alpha_e k_e^2  - q \sum_{v=1}^V  x_v^2 
+ 2 i ( p_f  \epsilon_{fv} x_v   +  k_e\epsilon_{ev} x_v )}
\label{paraphasereg}
\end{equation}
and to deduce the ordinary amplitudes from a limit $q \to 0$.

The last modification we perform is to attribute a different weight $q_v$ to each vertex regulator.
This is more natural from the point of view of universal polynomials.
So we define
\begin{equation}
B^T_G(p_1,...,p_N; \{q_v\} ) = \int  \prod_{e =1}^E  \big[
d \alpha_e e^{-\alpha_e m^2}   d^d k_e  \big]  \prod_{v=1}^{V} d^d x_v
e^{-\alpha_e k_e^2  - q_v x_v^2 
+ 2 i ( p_f  \epsilon_{fv} x_v   +  k_e\epsilon_{ev} x_v )}.
\label{paraphaseregs}
\end{equation}
These amplitudes are Gau\ss ian in the external variables $p_f$ and no longer involve
any noncanonical choice. We shall now compute their generalized Symanzik polynomials 
and deduce the ordinary Symanzik polynomials from these as 
leading terms when all $q_v$'s are sent to 0.

\subsection{Generalized Symanzik Polynomials}

\label{symanpoly}
We consider the phase space representation  (\ref{paraphaseregs}).
We have to perform a Gau\ss ian integral in $E+V$ variables (each of which is $d$-dimensional).
We consider these momentum and position variables as a single vector.
We also forget the label $^T$ for truncation as it is no longer needed in this section. The
graph we consider may be connected or not.

We introduce the condensed notations: 
\bea B_G (p_f, q_v) &=& \int   \prod_e
d\alpha_e    e^{-\alpha_e m^2}  d^d k_e  \int \prod_v d^d x_v 
e^{-  Y X_G Y^t} \label{mainformb}
\eea 
where $X_G$ is a $d(E+V+N)$ by   $d(E+V+N)$ square matrix, namely 
\begin{equation} X_G =
\begin{pmatrix}
\alpha_e &  - i \epsilon_{ev} &  0  \\
- i \epsilon_{ev}  & q_v &     - i \epsilon_{fv}  \\
0 &   - i \epsilon_{fv} & 0\\
\end{pmatrix}
\end{equation}
where $\alpha_e$ and $q_v$ are short notations for diagonal matrices
$\alpha_e \delta_{e,e'}$ and $q_v \delta_{v,v'}$. 
$Y$ is an $E+V+N$ by 1 line, namely
$Y = \begin{pmatrix}
k_e & x_v & p_f\\
\end{pmatrix} $.

We can further decompose $X_G$ as
\bea \label{mainformq}
X_G= \begin{pmatrix} Q_G &  - i R_G^{t} \\ - i R_G & 0 \\
\end{pmatrix}\ .
\eea
where $Q_G  =
\begin{pmatrix} \alpha_e &  - i \epsilon_{ev}  \\
- i \epsilon_{ev}  & q_v   \\ \end{pmatrix} $ is a  $d(E + V)$ by  $d(E + V)$ square matrix and 
$R_G$ is the real rectangular $N$ by $E+V$ matrix
made of a $dN$ by $dE$ zero block and the $dN$ by $dV$ "incidence flag" matrix $ \epsilon^{\mu}_{fv}$.
The dimensional indices $\mu$ being quite trivial we no longer write them down from now on.

Note $P$ the line $p_f$, hence the last part of the line $Y$.
Gau\ss ian integrations can be performed explicitly and the result
is a Gau\ss ian in external variables. Therefore up to inessential constants 
\bea B_G (p_f, q_v) &=& \int \prod_e d\alpha_e e^{-\alpha_e m^2}
\frac{1}{\det Q_G^{d/2}}    e^{ -  P R_G Q_G^{-1}  R_G^{t} P^t  }
\nonumber
\\ &=& \int   \prod_e d\alpha_e    e^{-\alpha_e m^2}  d^d k_e  
\frac {e^{-  {\cal V} / {\cal U} }}{{\cal U}^{d/2}}
\label{defampliB}
\eea
for some polynomial ${\cal U}_G$ in $\alpha$'s and $q$'s and 
a quadratic form in the $p$ variable ${\cal V}_G$ with polynomial
coefficients in $\alpha$'s and $q$'s.

\begin{definition}
The generalized Symanzik polynomials with harmonic regulators are the polynomials
appearing in (\ref{defampliB}), namely
\be
{\cal U}_G (\alpha_e, q_v)  = \det Q_G,
\ee
\be \label{secondpol}
{\cal V}_G (\alpha_e, q_v,p_f)/  {\cal U}_G (\alpha_e, q_v)   = P R_G Q_G^{-1}  R_G^{t} P^t .
\ee
\end{definition}

These polynomials can be computed explicitly:
\begin{theorem}\label{symantheo}
\be
{\cal U}_G (\alpha_e, q_v)  = \sum_{{\cal F}}  \prod_{e \not \in {\cal F}} \alpha_e  \prod_{{\cal C}} q_{\cal C}  ,
\ee
\be
{\cal V}_G (\alpha_e, q_v,p_f) = \sum_{{\cal F}}  \prod_{e \not \in {\cal F}} \alpha_e  
\sum_{{\cal C}}  p_{\cal C}^2   \prod_{{\cal C}' \ne {\cal C}} q_{{\cal C}'}  ,
\ee
where the sum over ${\cal F}$ runs over all forests of the graph, and the indices ${\cal C}$ and ${\cal C}'$
means any connected component of that forest (including isolated vertices if any).
The variables $p_{{\cal C}}$ and $q_{{\cal C}}$ are the natural sums associated
to these connected components.
\end{theorem}

In order to prove this theorem we introduce now the 
quasi-Grassmann representations of ${\cal U}_G$
and ${\cal V}_G$ of Lemma \ref{quasipfaff}.

Let's calculate first ${\cal U}$, hence the determinant of $Q_G$. 
Factoring out powers of $i$ we get:
\begin{equation}
\det Q_G = \det 
\begin{pmatrix}
\alpha_e&- \epsilon_{e v}\\
\epsilon_{e v}&q_v\\
\end{pmatrix}\\
\end{equation}
which can be written as sum of a diagonal matrix $D$, with 
coefficients $D_{ee} =\alpha_e$ and $D_{vv} = q_v$
and of an antisymmetric matrix $A$ with elements
$\epsilon_{ev}$, that is,  $Q=D+A$.

By Lemma \ref{quasipfaff}
\begin{eqnarray}
{\cal U}_G (\alpha_e, q_v) = \int \prod_{v,e}  d\chi_v d\omega_v  d\chi_e d\omega_e
e^{ - \alpha_e\chi_e \omega_e}e^{ -q_v\chi_v \omega_v}  e^{- \chi_e
\epsilon_{ev} \chi_v  + \omega_e
\epsilon_{ev} \omega_v } .
\end{eqnarray}

Similarly ${\cal V}$ which is a minor related to the $Q_G$ matrix is given by a Grassmann
integral but with sources
\begin{eqnarray}
{\cal V}_G (\alpha_e, q_v,p_f) &=&  \int \prod_{v,e}  d\chi_v d\omega_v  d\chi_e d\omega_e 
e^{ -\alpha_e\chi_e\omega_e
}e^{- q_v\chi_v\omega_v}  e^{-\chi_e
\epsilon_{ev}\chi_v   + \omega_e \epsilon_{ev}\omega_v    }\nonumber
\\&&  p_f \cdot p_{f'}   \epsilon_{fv}  \epsilon_{f'v'}   (\chi_v \omega_{v'}  + \chi_{v'} \omega_{v} ) ,
\end{eqnarray}
where we have expanded $ \bar \psi_v   \psi_{v'}  $ as $\frac{1}{2}  [\chi_v \chi_{v'}  
+ \omega_v \omega_{v'} + i (\chi_v \omega_{v'}  + \chi_{v'} \omega_{v} )]$ and canceled out the 
$\chi_v \chi_{v'}   + \omega_v \omega_{v'} $ term which must vanish by symmetry and the $i$
factors.

Now we can prove directly that these polynomials obey a deletion-contraction rule.
\begin{theorem}\label{grasstheo}
For any semi-regular edge $e$
\bea \label{delcontr1}
 {\cal U}_G (\alpha_e, q_v) = \alpha_e \;  {\cal U}_{G-e} (\alpha_e, q_v) + {\cal U}_{G/e} (\alpha_e, q_v) ,
\eea
\bea \label{delcontr2}
 {\cal V}_G (\alpha_e, q_v, p_f) = \alpha_e {\cal V}_{G-e} (\alpha_e, q_v, p_f ) + {\cal V}_{G/e} (\alpha_e, q_v,p_f).
\eea
Moreover we have the terminal form evaluation 
\be\label{firsttermin} {\cal U}_G (\alpha_e, q_v)  = \prod_{e}  \alpha_e   \prod_v q_v ,
\ee
\be\label{secondtermin} {\cal V}_G (\alpha_e, q_v, p_f)  = \prod_{e}  \alpha_e   \sum_v p_v^2
\prod_{v'\ne v} q_v
\ee
for $G$ solely made of self-loops attached to isolated vertices.
\end{theorem}
\prf  
If $G$ is not a terminal form we can pick up any semi-regular edge $e$ connecting vertices $v_1$ and 
$v_2$ with $\epsilon_{v_1} = +1, \epsilon_{v_2} = -1$. We expand
\be e^{- \alpha_e\chi_e \omega_e} = 1 + \alpha_e\omega_e\chi_e . 
\ee
For the first term, since we must saturate the $\chi_e$ 
and $\omega_e$ integrations, we must keep the $\chi_e (\chi_{v_1} - \chi_{v_{2}})  $ term in $e^{\sum_{v}
\chi_e \epsilon_{ev}\chi_v }$ and the similar $\omega$ term, hence we get a contribution
\begin{eqnarray}
\det Q_{G,e,1}&=& \int  \prod_{e'\ne e,v}  d \chi_{e'}d\omega_{e'}  d\chi_v d\omega_v   
(\chi_{v_1} - \chi_{v_2})  (\omega_{v_1} - \omega_{v_2}) 
\nonumber\\
&&e^{-\sum_{e'\ne  e }\alpha_e'\chi_{e'}
\omega_{e'}} e^{- q_v\chi_v\omega_v} e^{-\sum_{e' \ne e ,v}   \chi_{e'}
\epsilon_{e'v}\chi_v+ \sum_{e' \ne e ,v}   \omega_{e'}
\epsilon_{e'v}\omega_v   } .
\end{eqnarray}
Performing the trivial triangular  change of variables  with unit Jacobian:
\be  \hat \chi_{v_1} = \chi_{v_1} - \chi_{v_2}, \ \   \hat \chi_{v} = \chi_{v}\ \   \emph{for } \ v \ne v_1,
\ee
and the same change for the $\omega$ variables
we see that the effect of the $(\chi_{v_1} - \chi_{v_2})  (\omega_{v_1} - \omega_{v_2})$ term 
is simply to change the $v_1$ label into $v_2$ and to destroy the edge $e$ and the vertex $v_1$.
This is exactly the contraction rule, so $ \det Q_{G,e,1} =  \det Q_{G/e}$. The second term
$\det Q_{G,e,2}$ with the $\alpha_e\omega_e\chi_e$
factor is even easier. We must simply put to 0 all terms involving the $e$ label, hence trivially
$ \det Q_{G,e,2} = \alpha_e \det Q_{G-e}$.  Remark that during the contraction steps the
weight factor $q_{v_1}  \chi_{v_1} \omega_{v_1}  $ is just changed into  $q_{v_1}  \chi_{v_2} \omega_{v_2}  $.
That's why we get the new weight $q_{v_1} + q_{v_2}$ for the new vertex $v_2$ which represent
the collapse of former vertices $v_1$ and $v_2$.

Note that the source terms in ${\cal V}$ do not involve $\chi_e$ and $\omega_e$ variables.
Therefore the argument goes through exactly in the same way for the second polynomials. 
The only remark to make is that like weights, flag momenta follow contraction moves.

The evaluation on terminal forms is easy. For a graph with only vertices and self-loops the matrix $Q_G$
is diagonal, because $\epsilon_{ev}$ is always 0. 
Hence ${\cal U_G}$ is the product of the diagonal elements $\prod_e \alpha_e \prod_v q_v$.
The second polynomial can be analyzed through the Grassmann representation, but it is simpler to 
use directly (\ref{secondpol}) and the fact that $Q_G$ is diagonal to get (\ref{secondtermin}). This completes
the proof of Theorem \ref{grasstheo}, hence also of Theorem \ref{symantheo}.
\qed

We turn now to the limit of small regulators $q_v$ to show how for a connected graph $G$
the ordinary amplitude $ \delta (\sum_f p_f) A_G$ and the ordinary polynomials $U_G$ and $V_G$ emerge 
out of the leading terms of the regularized amplitude $B_G$
and the generalized polynomials ${\cal U}_G$ and ${\cal V}_G$.

When all $q$'s are sent to zero there is no constant term in ${\cal U}_G$ but a constant 
term in ${\cal V}_G$. Up to second order in the $q$ variables we have:
\be
{\cal U}_G (\alpha_e, q_v)  = q_G\sum_{{\cal T}}  \prod_{e \not \in {\cal T}} \alpha_e + O(q^2) ,
\ee
\be
{\cal V}_G (\alpha_e, q_v,p_f) =p^2_G\sum_{{\cal T}}  \prod_{e \not \in {\cal T}} \alpha_e +
 \sum_{{\cal T}_2} (p_{G_1}^2  q_{G_2} +  p_{G_2}^2 q_{G_1} )  \prod_{e \not \in {\cal T}_2} \alpha_e  
 + O(q^2) ,
\ee
where the sum over ${\cal T}$ runs over trees and the sum over ${\cal T}_2$ runs over 
two trees separating the graph into two connected components $G_1$ and $G_2$.
Hence we find 
\be \frac{  e^{- {\cal V} / {\cal U} }  }{{\cal U}^{d/2}}  = \frac{e^{- p^2_G / q_G }}{q_G^{d/2}} \frac{e^{  -
 \sum_{{\cal T}_2}(p_{G_1}^2  q_{G_2} +  p_{G_2}^2 q_{G_1} )  \prod_{e \not \in {\cal T}_2} \alpha_e  
/  q_G\sum_{{\cal T}}  \prod_{e \not \in {\cal T}} \alpha_e  +  p^2_G O(1)  
+ O(q)} }{ [ \sum_{{\cal T}}  \prod_{e \not \in {\cal T}} \alpha_e  + O(q)  ]^{d/2} } .
\ee
Up to inessential normalization factors the first term tends to $\delta(p_G)$ and the second one tends to $e^{-V/U}/U^{d/2}$
if we use the fact that $\delta(p_G) f(p_G) = \delta(p_G) f(0) $, that is if we
use the delta distribution to cancel the $p^2_G O(1)$ term and to 
simplify $(p_{G_1}^2  q_{G_2} +  p_{G_2}^2 q_{G_1} ) $
into $q_G p^2_{G_1} = q_G p^2_{G_2} $. This proves (\ref{limBA}).

The $U_G$ and $V_G$ polynomials 
are in fact easy to recover simply from the ${\cal U}_G$ polynomial alone:
\begin{theorem} For any connected $G$ and any vertex $v$
\be
U_G (\alpha_e) =\frac { \partial }{\partial q_v} {\cal U}_G (\alpha_e, q_v)  \,\,\;    \vert_{q_{v'} = 0 \ \forall v'} .
\ee
On the submanifold $p_G =0$ we further have
\be
V_G (\alpha_e, p_f) = - \frac{1}{2} \sum_{v \ne v'}  p_v \cdot p_{v'} \,\,\; 
 \frac{ \partial^2 }{\partial q_v\partial q_{v'}  } �{\cal U}_G (\alpha_e, q_v) \,\,\;  \vert_{q_{v"} = 0 \  \forall v"} .
\ee
\end{theorem}
\prf It is an easy consequence of Theorem \ref{symantheo}. 
\qed

We can also prove an analog of Proposition \ref{propUV} between
${\cal V}_G$ and ${\cal U}_{G(vv')}$ but only on the submanifold $p_G =0$.

\subsection{Relation to discrete Schr\"odinger Operator}

As an aside, it is worthwhile to notice that there is a relation with discrete Schr\"odinger operators on graphs \cite{schrodinger}. Recall that given a graph $G=(V,E)$, the discrete Laplacian is defined as follows. We first introduce the 0-forms $\Omega_{0}(G)={\mathbb R}^{V}$ as the real functions on the set of vertices and 1-forms $\Omega_{0}(G)={\mathbb R}^{E}$ as functions on the edges. Then, the discrete differential $\mathrm{d}:\,\Omega_{0}(G)\rightarrow\Omega_{1}(G)$ is defined as
\begin{equation}
\mathrm{d}\psi(e)=\sum_{v}\epsilon_{ev}\,\psi_{v},
\end{equation}
where we recall the convention that for a self-loop $\epsilon_{e v}=0$ and an arbitrary orientation is chosen on the edges. Next, given strictly positive weights $\beta_{e}$ associated to the edges, we define $\mathrm{d}^{\ast}:\,\Omega_{1}(G)\rightarrow\Omega_{0}(G)$ by
\begin{equation}
\mathrm{d}^{\ast}\!\phi(v)=\sum_{e}\beta_{e}\epsilon_{ev}\,\phi_{e}.
\end{equation}
Note that $\mathrm{d}^{\ast}$ is precisely the adjoint of $\mathrm{d}$ for the scalar product on ${\mathbb R}^{E}$ defined by the weights $\beta_{e}$ and the Euclidean one on ${\mathbb R}^{V}$. Accordingly, the 0-form Laplacian $\Delta:\,\Omega_{0}(G)\rightarrow\Omega_{0}(G)$ is
\begin{equation}
\Delta=\mathrm{d}^{\ast}\mathrm{d},
\end{equation}
or, in terms of its action on functions $\psi\in{\mathbb R}^{V}$,
\begin{equation}
\Delta\psi(v')=\sum_{e,v}\beta_{e}\epsilon_{ev'}\epsilon_{ev}\,\psi_{v}.
\end{equation}
Note that there is exactly one zero mode per connected component, as follows from the equivalence between $\Delta\psi=0$ and $\mathrm{d}\psi=0$. Finally, the weights $q_{v}$ associated to the vertices\footnote{Strictly speaking, the latter are associated to the flags and $q_{v}$ is the sum the weights of the flags attached to $v$.} define a function $V$ from the vertices to ${\mathbb R}$ acting multiplicatively on $\Omega_{0}(G)$  so that we define the discrete Schr\"odinger operator (Hamiltonian in the quantum mechanics language) on the graph by
\begin{equation}
H=-\Delta+V.
\end{equation}
Turning back to the parametric representation, if we perform the Gau\ss ian integration over the momenta we are left with
\begin{equation}
\frac{\pi^{D/2}}{(\alpha_{1}\cdots\alpha_{e})^{D/2}}\int {\textstyle \prod_{v}dx_{v}}\,\mathrm{e}^{-\sum_{v,v'}x_{v}H_{v,v'}x_{v'}+2\mathrm{i}\sum_{v}x_{v}\cdot p_{v}},
\end{equation}
with weights $\beta_{e}=\frac{1}{\alpha_{e}}$. In particular, the first Symanzik polynomial with regulators $q_{v}$ is expressed in terms of the determinant of $H$,
\begin{equation}
{\cal U}_{G}(\alpha,q)=\left({\textstyle \prod_{e}\alpha_{e}}\right)\,\det H=
\left({\textstyle \prod_{e}\alpha_{e}}\right)\int {\textstyle \prod_{v}d\,\overline{\psi}_{v}d\psi_{v}}\,
\mathrm{e}^{-\sum_{v,v'}\overline{\psi}_{v}H_{v,v'}\psi_{v'}},
\end{equation}
with $\overline{\psi}_{v},\psi_{v}$ Grassmann variables. By the same token, the ratio appearing in the Feynman amplitude is expressed in terms of its inverse $G$ (Green's function in the quantum mechanics language), 
\begin{equation}
\frac{{\cal V}_{G}(\alpha,q,p)}{{\cal U}_{G}(\alpha,q)}=
\sum_{v,v'}G_{v,v'}\,p_{v}\cdot p_{v},
\end{equation}
where the Green's function can also be expressed using Grassmann integrals. As a byproduct, it turns out that it can also be computed by contraction/deletion.

\subsection{Categorified Polynomials}

We have up to now considered two seemingly unrelated graph polynomials obeying contraction/deletion rules, the multivariate Tutte polynomial  $Z_{G}(\beta_{e},q)$ and ${\cal U}_{G}(\alpha_{e},q_{i})$, from which the Symanzik polynomials can be recovered by various truncations. Therefore, it is natural to wonder wether there is a single graph polynomial, obeying contraction/deletion rules too, from which both $Z_{G}(\beta_{e},q)$ and ${\cal U}_{G}(\alpha_{e},q_{i})$ can be recovered. In this subsection for simplicity we shall consider only
the first Symanzik polynomial, and the flags considered in this subsection no longer bear external momenta,
but an abstract index.

Such a polynomial is an invariant of graphs with flags, i.e. labeled half-edges attached to the vertices. 
In order to make the contraction possible, it is necessary to allow each vertex to have several flags, all carrying distinct labels. The requested polynomial, ${\cal W}_{G}(\beta_{e},q_{I})$ depends on edge variables $\beta_{e}$ as well as on independent variables $q_{I}$ for each non empty subset $I$ of the set of labels of the flags,  with the proviso that, for each vertex, the subsets $I$ contain all the flags attached to the vertex or none of them. Thus, for a diagram with $V'$ vertices carrying flags there are $2^{V'}-1$ variables $q_{I}$. 

\begin{definition}
For a graph $G$ with flags, ${\cal W}_{G}(\beta_{e},q_{I})$  is defined by the expansion
\begin{equation}
{\cal W}_{G}(\beta_{e},q_{I})=\sum_{A\subset E}\,\Big(\prod_{e\in E}\beta_{e}
\hspace{-0.5cm}\prod_{{\cal C}_{n}\atop
\mbox{\tiny connected components}}\hspace{-0.5cm}q_{I_{n}}\Big),
\end{equation}
where $I_{n}$ are the sets of flags attached to the vertices of the connected component ${\cal C}_{n}$ of the spanning graph $(V,A)$. 
\end{definition}

For example, for the bubble graph on two vertices with 
two edges between these vertices and flags $1,2$ attached to one of 
vertex, and flag 3 to the other one, we have
\begin{equation}
{\cal W}_{G}(\beta_{e},q_{I})=(\beta_{1}\beta_{2}+\beta_{1}+\beta_{2})q_{123}
+q_{12}q_{3}.
\end{equation}

Since the variables $q_{I}$ are defined using the flags, the contraction/deletion rule for ${\cal W}_{G}(\beta_{e},q_{I})$ requires us to properly define how the flags follow the contraction/deletion rule for any edge of $G-e$ and $G/e$. Because the vertices and the flags of $G-e$ are left unchanged, the same variables $q_{I}$ appear in $G$ and $G-e$. For $G/e$, we restrict the $q_{I}$ to those associated with subsets that contain either all the flags attached to the two vertices merged by the contraction of $e$, either none of them. This is best formulated using flags: the new vertex simply carries the flags of the two vertices that have been merged. Then, the contraction/deletion identity simply follows from grouping the terms in ${\cal W}_{G}(\beta_{e},q_{I})$ that contain $\beta_{e}$ and those that do not.

\begin{proposition}
The polynomial ${\cal W}_{G}(\beta_{e},q_{I})$ obeys the contraction/deletion rule for any edge
\begin{equation}
{\cal W}_{G}(\beta_{e},q_{I})=\beta_{e}{\cal W}_{G/e}(\beta_{e'\neq e},q_{I}|_{G/e})+{\cal W}_{G-e}(\beta_{e'\neq e},q_{I}).
\end{equation}
\end{proposition}

The multivariate Tutte polynomial is easily recovered by setting $q_{I}=q$ for any $I$,
\begin{equation}
Z_{G}(\beta_{e},q)={\cal W}_{G}(\beta_{e},q_{I}\!=\!q).
\end{equation}
In this case, all the information about the flags is erased and so that the latter may be omitted. To recover ${\cal U}_{G}(\alpha_{e},q_{i})$, it is convenient to introduce as an intermediate step the polynomial
\begin{equation}
{\Upsilon}_{G}(\alpha_{e},q_{i})=\sum_{A\subset E}\,\prod_{e\notin E}\alpha_{e}\prod_{{\cal C}_{n}}\Big(\sum_{i\in I_{n}}q_{i}\Big),
\end{equation}
where as before $I_{n}$ are the flags included in the connected component ${\cal C}_{n}$ of the spanning graph $(V,A)$. By its very definition, ${\Upsilon}_{G}(\alpha_{e},q_{i})$ is related to ${\cal W}_{G}(\beta_{e},q_{I})$ by setting $q_{I}=\sum_{i\in I}q_{i}$,
\begin{equation}
{\Upsilon}_{G}(\alpha_{e},q_{i})=
\Big({\prod_{e}\alpha_{e}}\Big)\,{\cal W}_{G}(\beta_{e}\!=\!1/\alpha_{e},q_{I}\!=\!{\textstyle \sum_{i\in I}}q_{i}).
\end{equation}
Then, the polynomial ${\cal U}_{G}(\alpha_{e},q_{i})$ is obtained from ${\Upsilon}_{G}(\alpha_{e},q_{i})$ by keeping only the highest degree terms in the $\alpha_{e}$'s for each term in $\prod_{{\cal C}_{n}}\sum_{i\in I_{n}}q_{i}$. Indeed, ${\cal U}_{G}(\alpha_{e},q_{i})$ is obtained from ${\Upsilon}_{G}(\alpha_{e},q_{i})$ by truncating its expansion to those subsets $A\subset E$ that are spanning forests, i.e. that obey $0=|A|-V+k(A)$. Since the number of connected components $k(A)$ is fixed by the global degree in the $q_{i}$'s, the forests are obtained with $|A|$ minimal, so that the global degree in the $\alpha_{e}$'s must be maximal. Note that a truncation to the spanning forests may also be performed at the level of the multivariate Tutte polynomial by restricting, at fixed degree in $q$, to the terms of minimal degree in the $\beta_{e}$'s. This yields an expansion over spanning forests \cite{Sokal1} (see also \cite{Sokal2}).
\begin{equation}
F_{G}(\beta_{e},q)=\sum_{A\subset E\atop \mbox{\tiny spanning forest}}\Big(\prod_{e\in A}\beta_{e}\Big)\, q^{k(A)}.
\end{equation}
This, as well as the relation to the Symanzik polynomial, is conveniently summarized by the following diagram.

\begin{proposition}
The previous polynomials may be obtained from ${\cal W}(\alpha_{e},q_{I})$ by the following series of substitutions and truncations,
\begin{equation}
\xymatrix{
&{\Upsilon}_{G}(\alpha_{e},q_{i})\ar[r]^{\mbox{\tiny highest order}\atop\mbox{\tiny in the }\,\alpha_{e}}&{\cal U}_{G}(\alpha_{e},q_{i})\ar[dr]^{\mbox{\tiny\quad  term in}\,\sum_{i}q_{i}}&\\
{\cal W}_{G}(\beta_{e},q_{I})\ar[ur]^{q_{I}=\sum_{i\in I}q_{i}\atop \mbox{\tiny \!\!multiplication by}\,\prod_{e}\!\alpha_{e}\quad\quad}\ar[dr]_{q_{I}=q}\quad&&&
U_{G}(\alpha_{e})\\
&
Z_{G}(\beta_{e},q)\ar[r]^{\mbox{\tiny lowest order}\atop\mbox{\tiny in the }\,\beta_{e}}&\ar[ur]_{\mbox{\tiny \quad term in}\,q\atop \mbox{\tiny multiplication by}\,\prod_{e}\!\alpha_{e}}{F}_{G}(\beta_{e},q)&
}
\end{equation}
where $\alpha_{e}=1/\beta_{e}$.
\end{proposition}

Alternatively, the polynomial ${\cal W}_{G}(\alpha_{e},q_{I})$ can be seen as an extension of the polynomial ${W}_{G}(\xi_{a},y)$ introduced by Noble and Welsh in \cite{Wpolynomial}. 

\begin{definition}
For a graph with weights $\omega_{v}\in{\mathbb N}^{\ast}$ assigned to the vertices, the $W$ polynomial is defined as 
\begin{equation}
{W}_{G}(\xi_{a},y)=\sum_{A\subset E}\hspace {0.5cm}(y-1)^{|A|-r(A)}\hspace{-1cm}\prod_{{\cal C}_{1},\dots,{\cal C}_{k(A)}\atop\mbox{\tiny connected components of} \, (V,A)}\hspace{-1cm}\xi_{a_{n}}
\end{equation}
with $a_{n}=\sum_{v\in {\cal C}_{n}}\omega_{v}$ the sum of the weights of the vertices in the connected component ${\cal C}_{n}$. 
\end{definition}

This polynomial also obeys the contraction/deletion rule if we add the weights of the two vertices that are merged after the contraction of an edge. Alternatively, weights may be assigned to flags, with the convention that the weight of a vertex is the sum of the weights of the flags attached to it. Then, ${W}(\xi_{a},y)$ is naturally extended to diagrams with flags  and results from a simple substitution in ${\cal W}_{G}(\xi_{a},y)$.

\begin{proposition}
For a graph with weights $\omega_{i}\in{\mathbb N}^{\ast}$ assigned to the flags,  
\begin{equation}
{W}_{G}(\xi_{a},y)=(y-1)^{-|V|}{\cal W}_{G}\big(\beta_{e}\!=\!y\!-\!1,q_{I}\!=\!(y-\!\!1)\xi_{a_{I}}\big),
\end{equation}
with $a_{I}=\sum_{i\in I}\omega_{i}$ the sum of the weights of the flags in $I$.
\end{proposition}
  
The polynomial $W_{G}(\xi_{a},y)$ only encodes the sum of the weights of the flags in each connected component and erases information about their labels. In particular, if we weight each flag by $\omega_{i}=1$, then the expansion of $W$ only counts the number flags per component whereas that of  ${\cal W}_{G}(\beta_{e},q_{I})$ keeps track of the associated set of labels. In a more sophisticated language, the latter may be considered as the simplest categorification of the former: integers, understood as finite sets up to isomorphisms, have been replaced by the category of finite sets.

\subsection{Symanzik Polynomials through the tree matrix theorem in $x$-space}

In this section we provide a sketch of a more standard proof of Theorem \ref{theosym} through the 
$x$ space representation and the tree matrix theorem.
The reason we include it here is for completeness and because we have not been able to find it in the 
existing literature, in which the same computation is usually performed through the
Binet-Cauchy theorem.

The $V \times V$ matrix $Q_G(\alpha)$ analog in this case of (\ref{mainformq}) is defined as
\begin{equation}
[Q_G(\alpha)]_{v,v'}=\sum_e \epsilon_{ev}\frac{1}{\alpha_e}\epsilon_{ev'}.
\end{equation}
It has vanishing sum over  lines (or columns):
\be \sum_{v'}  [Q_G(\alpha)]_{v,v'}=\sum_{v'} \sum_e\epsilon_{ve }\frac{1}{\alpha_e}\epsilon_{ev'} = 0 .
\ee

Therefore by the tree matrix theorem the determinant 
of the $(V-1)\times(V-1)$ matrix $Q_G(\alpha)$ defined as its principal minor
with the line and column for the root vertex number $V$ deleted is:
\begin{equation}\Delta_G(\alpha)=\det[Q_G(\alpha)]=
\sum_{{\cal T}}\prod_{e\in{\cal T}}\frac{1}{\alpha_e}
\end{equation} where the sum is over all trees of G. Since every tree of G has
$V-1$ edges, $\Delta_G$ is clearly a homogenous polynomial in the
$\alpha_e^{-1}$.  For $\alpha>0$, $\Delta$ is positive. The remaining
$(V-1)$ vectors $z$ may then be integrated over and the result
is
\begin{equation}
A_{G}(p)=\int_0^\infty\prod_l(d\alpha_e e^{-\alpha_e
m^2})\frac{\exp\{ - p_{v}
[Q^{-1}_G(\alpha)]_{v,v'}p_{v'}\} }{[\alpha_1...\alpha_E \Delta_G(\alpha)]^{d/2}} .
\end{equation}

This formula expresses $A_G(p)$ as a function of the invariant
scalar product of external momenta $p_{v} \cdot p_{v'}$.
The denominator 
\begin{equation}
 U_G(\alpha)\equiv
\alpha_1...\alpha_E\Delta_G(\alpha)=\sum_{{\cal T}}\prod_{e\not\in{\cal T}}\alpha_e
\end{equation}
is a homogenous polynomial of degree $V-1$. This gives an alternative 
proof of (\ref{paramet}). The second Symanzik polynomial can also be obtained
through this method and the corresponding computation is left to the reader. 
Of course harmonic regulators can also be included
if one wants to avoid the noncanonical choice of a root, but the Pfaffian structure
of the phase space representation is lost. Also this $x$-space method does not generalize easily
to noncommutative field theory to which we now turn our attention.

\section{Bollob\'as-Riordan Polynomials}

\label{briordan}
\subsection{Ribbon graphs}

\begin{figure}
\begin{center}
\includegraphics[scale=0.9,angle=-90]{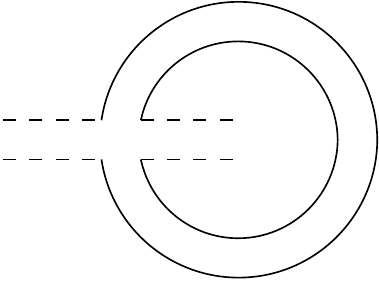}
\caption{A planar ribbon graph with $V=E=1$. $bc=2$ and two flags.}\label{figpla}
\end{center}
\end{figure}

\begin{figure}
\begin{center}
\includegraphics[scale=1]{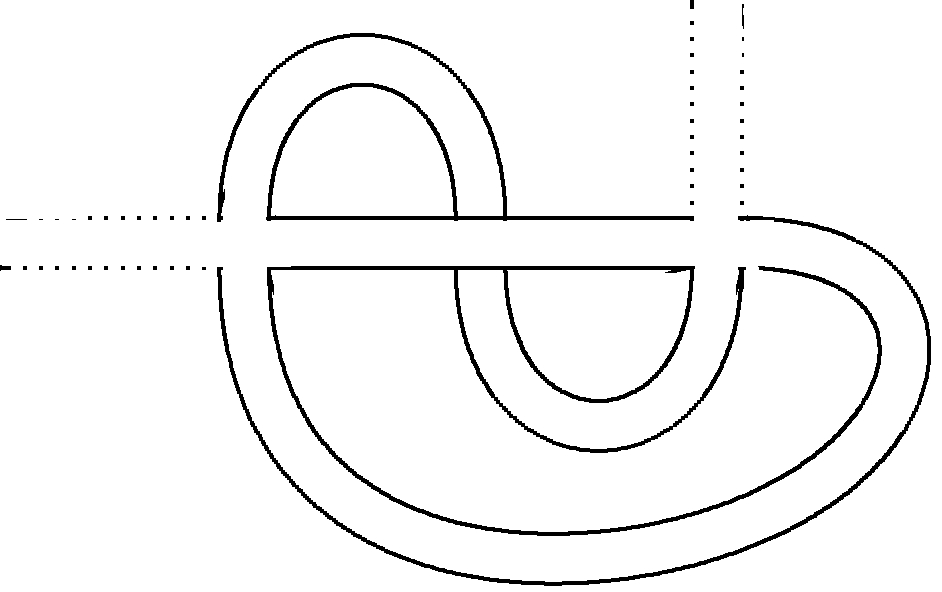} \hskip2cm \includegraphics[scale=1]{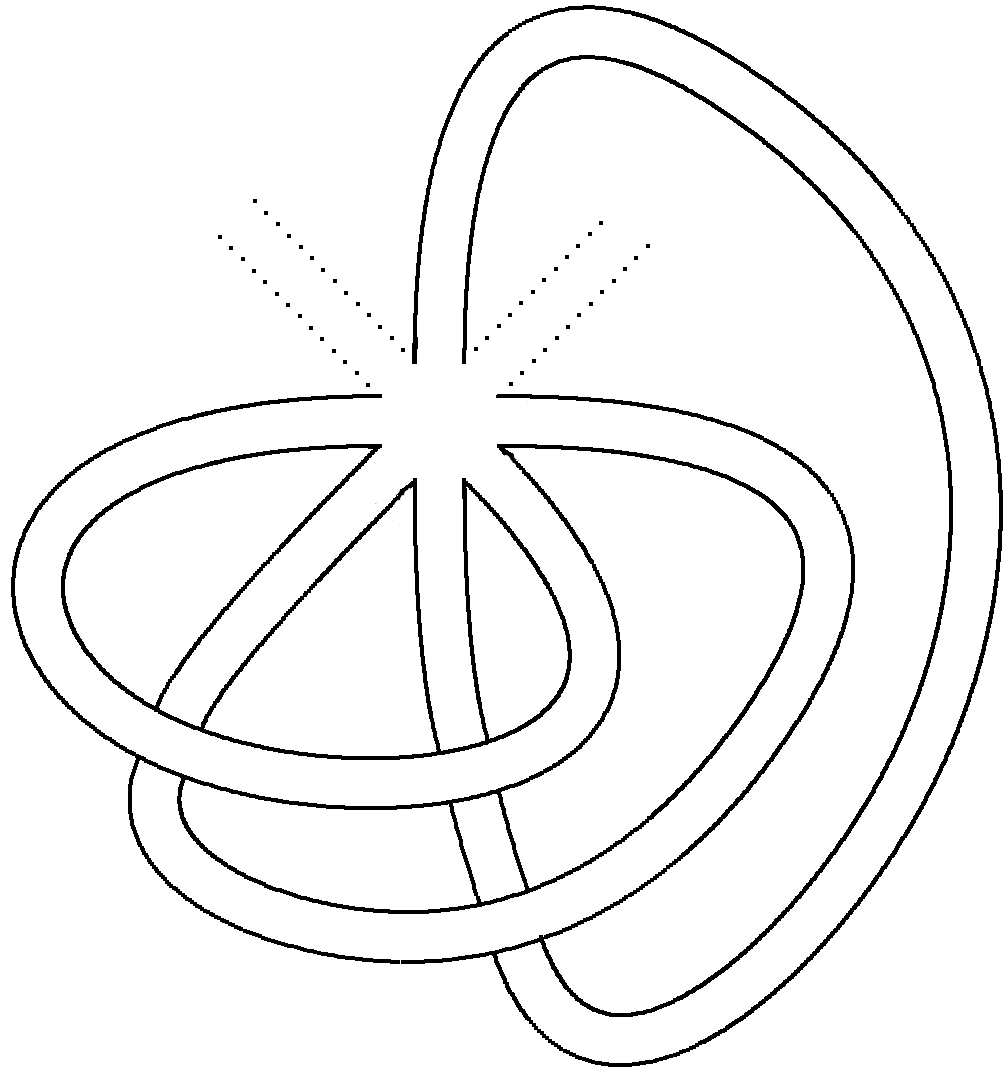}
\caption{A non-planar ribbon graph without flags, with $V=2$, $E=3$, $bc=1$, $g=1$, $f=2$,
and its dual graph with $V=1$, $E=3$, $bc=2$, $g=1$, $f=2$.}\label{figdual}
\end{center}
\end{figure}

A ribbon graph $G=(V,E)$ is an orientable surface with
boundary represented as the union of $V$ closed disks, also called
vertices, and $E$ ribbons also called edges, such that:
\begin{itemize}
\item the disks and the ribbons intersect in disjoint line
segments,
\item each such line segment lies on the boundary of precisely one
disk and one ribbon,
\item every ribbon contains two such line segments.
\end{itemize}
So one can think of a ribbon graph as consisting of
disks (vertices) attached to each other by thin stripes (edges) glued to
their boundaries (see Figures \ref{figpla}-\ref{figdual}). For any such ribbon graph $G$ there is an underlying
ordinary graph $\bar G$ obtained by collapsing the disks to points and the ribbons to edges.

Two ribbon graphs are isomorphic if there is a
homeomorphism from one to the other mapping vertices to vertices and
edges to edges. A ribbon graph is a graph with a fixed cyclic
ordering of the incident half-edges at each of its vertices.

A face of a ribbon graph is a connected component of its boundary as a surface.
If we glue a disk along the boundary of each face we obtain a closed Riemann surface
whose genus is also called the genus of the graph.
The ribbon graph is called planar if that Riemann surface has genus zero. 

Generalized  ribbon graphs that can also incorporate Moebius strips and correspond to nonorientable surface
can be defined but will not be considered in this paper.

There is a duality on ribbon graphs  which preserves the genus but exchanges
faces and vertices, keeping the number of edges fixed. It simply considers the disks glued along faces
as the vertices of a dual graph and changes the ends of each ribbon into borders of the dual ribbon.

Extended categories of ribbon graphs with flags can be defined. Flags can be represented 
as ribbons bordered by dotted lines to distinguish them from ordinary edges (see Figures  
\ref{figpla}-\ref{figdual}).
Beware that the cyclic ordering of flags and half-edges at each vertex is very important
and must be respected under isomorphisms. The genus of an extended graph is defined as the genus of the graph obtained by removing the flags and closing the corresponding segments on their vertices. The number of broken faces is the number of faces which do contain at least one flag. It is an important 
notion in noncommutative field theory.

We define for any ribbon graph
\begin{itemize}
\item $V(G)$ as the number of vertices; 
\item $E(G)$, the number of edges,
\item $k(G)$, the number of connected components,
\item  $r(G)=V(G)-k(G)$, the rank of $G$,
\item $n(G)=E(G)-r(G)$, the nullity of $G$,
\item $bc(G)= F(G)$, the number of components
of the boundary of $G$\footnote{This is the number of \emph{faces} of $G$ when $G$ is connected.},
\item $g(G)= k - (V - E + bc)/2$ is the genus of the graph,
\item $f(G)$ the number of flags of the graph.
\end{itemize}

A graph with a single vertex hence with $V=1$ is called a \emph{rosette}.

A subgraph $H$ of a ribbon graph $G$ is a subset of the edges of $G$. 

The Bollob\'as-Riordan polynomial, which is a generalization of Tutte
polynomial, is a algebraic polynomial that is used to incorporate
new topological information specific to ribbon graphs, such as 
the genus and the number of "broken" or "external" faces. It is 
a polynomial invariant of the ribbon graph.


\subsection{Bollob\'as-Riordan Polynomial}

\begin{definition}[Global definition]
The Bollob\'as-Riordan polynomial is defined by:
\begin{eqnarray}
R_G= R_G(x,y,z)
=\sum_{H\subset
G}(x-1)^{r(G)-r(H)}y^{n(H)}z^{k(H)-bc(H)+n(H)}.
\end{eqnarray}
\end{definition}

The relation to the Tutte polynomial for the underlying graph $\bar G$
is $R_{G} (x-1, y-1  ,1) = T_{\bar G} (x,y)$. 
Remark also that if $G$ is planar we have $R_G (x-1,y-1,z) =  T_{\bar G} (x,y)$.

When $H$ is a spanning graph of $G$, we have
$k(H)-k(G)$=$r(G)-r(H)$. So we can rewrite the $R$ polynomial as:
\begin{equation}
R_G=  (x-1)^{-k(G)}\sum_{H\subset G} M(H),
\end{equation}
where
\begin{equation}\label{defbrior}
M(H)=(x-1)^{k(H)}y^{n(H)}z^{k(H)-bc(H)+n(H)}
\end{equation}
so that $M(H)$ depends only on $H$ but not on $G$.

\subsection{Deletion/contraction}\label{delcontractribbon}

\begin{figure}
\begin{center}
\includegraphics[scale=0.7,angle=0]{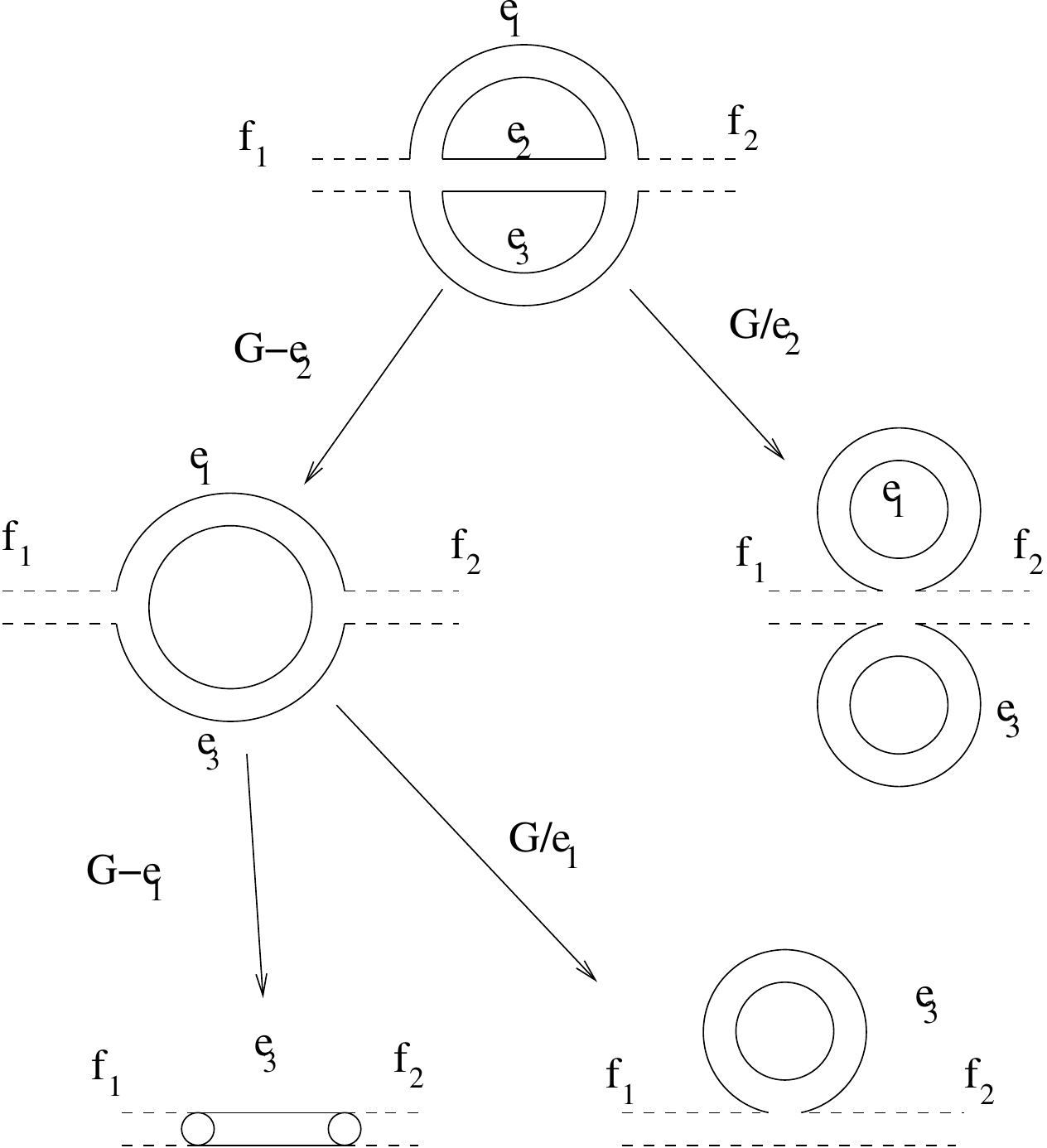}
\caption{The contraction-deletion for a ribbon graph.}\label{fig:contribbon}
\end{center}
\end{figure}

The deletion and contraction of edges in a ribbon graph are defined quite naturally:
the deletion removes the edge and closes the two scars at its end; the contraction
of a semi-regular edge
creates a new disk out of the two disks at both ends of the ribbon with a new
boundary which is the union of the boundaries of the two disks and of the ribbon (see 
Figure \ref{fig:contribbon}).
An interesting property is that deletion and contraction of edges  are exchanged in the dual graph. 

\begin{figure}
\begin{center}
\includegraphics[scale=0.5,angle=-90]{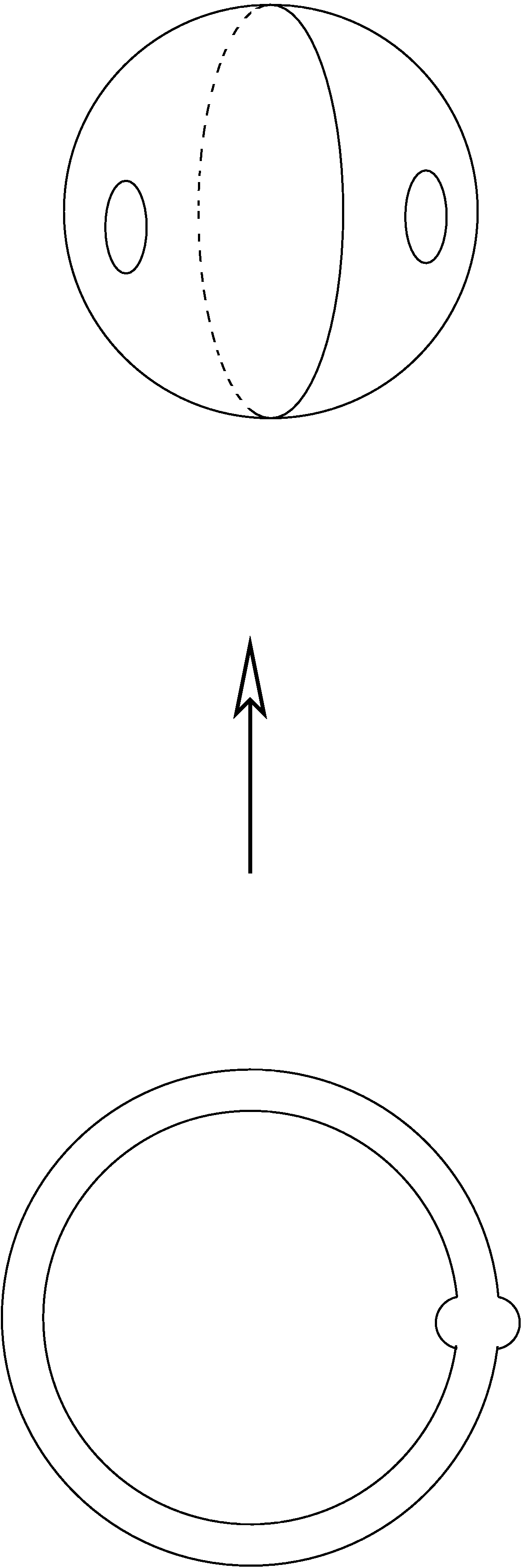}
\caption{Contraction of the single self-loop $G_1$.}\label{cyli}
\end{center}
\end{figure}

\begin{figure}
\begin{center}
\includegraphics[scale=0.5,angle=-90]{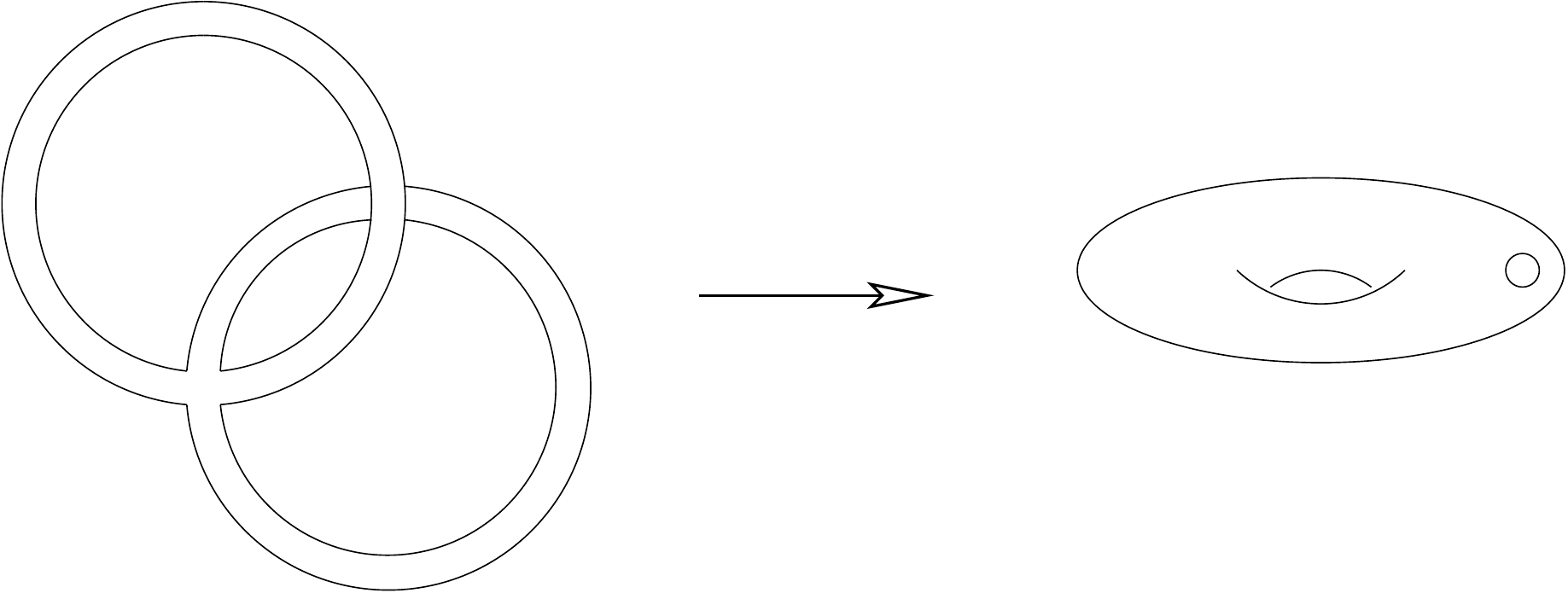}
\caption{Contraction of the two self loops non-planar $G_2$.}\label{torus}
\end{center}
\end{figure}

The deletion of a self-loop is standard. However the natural contraction of a self-loop creates
a surface with a new border. Iterating, we may get surfaces of arbitrary genus
with an arbitrary number of disks removed, a category also called 
disk-punctured surfaces.
The ribbons can now join any puncture to any other.
For instance the contraction of the self-loop on the graph $G_1$ of Figure  \ref{cyli} leads to a cylinder
ie to a single vertex which is a sphere with two disks removed.
The contraction of the two self-loops in graph $G_2$ of Figure \ref{torus} corresponds to the 
cylinder with a ribbon gluing
the two ends, hence to a torus.

Deletion and contraction defined in this extended category of graphs
can be iterated until the graph has no longer any edge,
ie is a collection of disk-punctured Riemann surfaces.
These punctured Riemann surfaces are very 
natural objects both in the context of string theory
and in NCQFT. However we do not consider them in this paper.

\begin{figure}
\begin{center}
\includegraphics[scale=0.85]{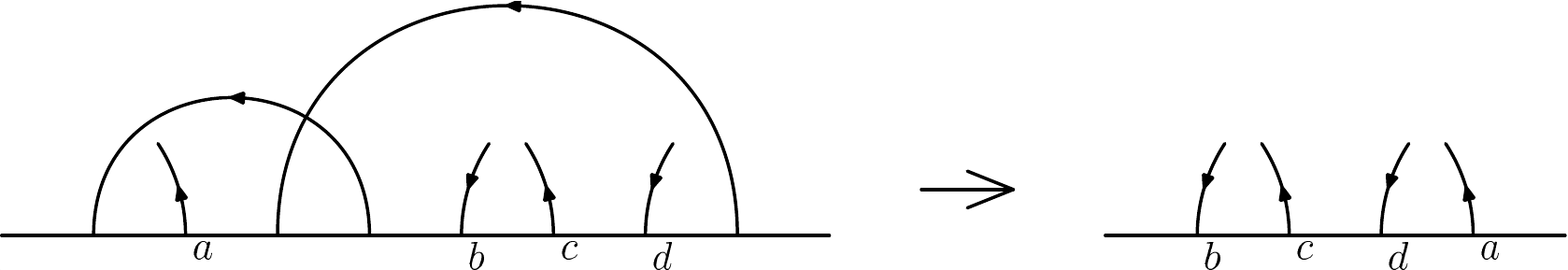}
\caption{When deleting the two edges of a nice pair crossing on some contracted vertex, 
one also needs to interchange the half-edges encompassed 
by the first edges with those encompassed by the second one. Beware
that the horizontal line in this picture is a part of the rosette cycle.}\label{filk3}
\end{center}
\end{figure}

In this paper we remain in the category of ordinary ribbon graphs with
disk-like vertices. The contraction/deletion of semi-regular edges
leads to rosettes as terminal forms. To treat them 
we introduce the notion of \emph{double contraction}
on \emph{nice crossings}.
Nice crossings were introduced in \cite{GurauRiv}:

\begin{definition}
A nice crossing pair of edges in a rosette is a pair of crossing edges $e_1$ and $e_2$
which are adjacent on the
cycle of the rosette. Adjacency means that one end of $e_1$ is consecutive with
an end of $e_2$ (see Figure \ref{filk3}).
\end{definition}
It is proved in \cite{GurauRiv} that any rosette ${\cal R}$ of genus $g>0$ contains at least one
nice crossing.

The double contraction of such a nice crossing pair consists
in deleting $e_1$ and $e_2$ and interchanging 
the half-edges encompassed by $e_1$ with the ones encompassed by $e_2$, 
see Figure \ref{filk3}.  This \emph{double contraction}
was defined in \cite{GurauRiv}
under the name of ``3rd Filk move''. It decreases the genus by one and 
the number of edges by 2.

In the next section iterating this double contraction until we reach
planarity allows us to compute the $U^\star $ Symanzik polynomial
by remaining in the category of ordinary ribbon graphs.

\begin{theorem}[Bollob\'as-Riordan polynomial, contraction/deletion]
\label{delcontrbollo}

\begin{equation}
R_G=R_{G/e}+R_{G-e}
\end{equation}
for every ribbon graph $G$ and any regular edge $e$ of $G$ and
\begin{equation}
R_G=x R_{G/e}
\end{equation}
for every bridge of $G$.
\end{theorem}

Therefore the R polynomial satisfy contraction-deletion relations 
as the Tutte polynomial. However to complete its definition 
we also need to define the R polynomial for single vertex graphs, 
namely the rosettes, which can be read off from (\ref{defbrior}).
For such a rosette ${\cal R}$, $k({\cal R})= V({\cal R}) = k(H) = V(H) =1$, so that the
R polynomial does not depend on $x$ and
\begin{equation}
R_{\cal R} (y,z)= \sum_{H \subset {\cal R}}  y^{E(H)} z^{2g(H)}.
\end{equation}
For $z=1$ we recover $R_{\cal R} (y-1,1) = y^{E(\cal R)}$.

\subsection{The multivariate Bollob\'as-Riordan polynomial}
Like in the case of Tutte polynomial, we can  generalize the
Bollob\'as-Riordan polynomial to a multivariate case. As before,we
associate to each edge $e$ a variable $\beta_e$. 

\begin{definition}
The multivariate Bollob\'as-Riordan polynomial of a
ribbon graph analog of the multivariate polynomial (\ref{multivartut}) is:
\begin{equation}
Z_G(x,\{\beta_e\},z)=\sum_{H\subset G} x^{k(H)}(\prod_{e\in H}  \beta_{e}) \, z^{bc(H)} .
\end{equation}
\end{definition}

It obeys again a deletion/contraction relation 
similar to Theorem (\ref{delcontrbollo}) for any semi-regular edge.

\section{Translation-invariant NCQFT}

\subsection{Motivation}

Noncommutative quantum field theory, hereafter called NCQFT, has a long story.
Schr\"odinger, Heisenberg  \cite{Schro} and Yang \cite{Yang} tried to extend the 
noncommutativity of phase space to ordinary space.
Building on their ideas Snyder \cite{Snyder} formulated
quantum field theory on such noncommutative space  in the hope that it 
might behave better than ordinary QFT in the ultraviolet regime.

Right from the start another motivation to study noncommutative quantum field theory
came from the study of particles in strong magnetic fields. It was early recognized
that non zero commutators occur for the coordinates of the centers
of motion of such quantum particles, so that noncommutative geometry 
of the Moyal type should be the proper setting for  many body quantum physics in strong external field.
This includes  in condensed matter the quantum Hall effect  
(see the contribution of Polychronakos in \cite{QS}), or other strong field situations.

An other motivation comes from particle physics.
After initial work by Dubois-Violette, Kerner and Madore,
Connes, Lott, Chamseddine and others have forcefully advocated 
that the \emph{classical} Lagrangian of the current standard
model arises naturally on a simple noncommutative geometry. For a review
see Alain Connes's contribution in  \cite{QS} and references therein.

Still an other motivation came from the search of new regularizations of non-Abelian gauge theories
that may throw light on their difficult mathematical structure. After
't~Hooft proposed the large $N$ limit of matrix theory, in which planar graphs dominate, as relevant
to the subject \cite{Hoo}, the Eguchi-Kawai model was an important attempt for an explicit solution.
These ideas have been revived in connection with 
the ultraviolet behavior of NCQFT on the Moyal-Weyl geometry, which also 
leads to the domination of planar graphs. Seiberg and Witten proposed in \cite{Seiberg1999vs} a mapping between ordinary and noncommutative gauge fields which does not preserve the gauge groups but preserve the gauge equivalent classes. 

The interest for non commutative geometry also stems from string theory.
Open string field theory may be recast as a problem of noncommutative
multiplication of string states \cite{Witten}.
It was realized in the late 90's that NCQFT is an effective theory of strings \cite{CDS}. 
Roughly this is because in addition to the symmetric tensor $g_{\mu\nu}$ the spectrum 
of the closed string also contains an antisymmetric tensor $B_{\mu\nu}$. There is no reason
for this antisymmetric tensor not to freeze at some lower scale into a classical field,
inducing an effective non commutative geometry of the Moyal type. 
There might therefore be some intermediate regime
between QFT and string theory where NCQFT is the relevant formalism. The ribbon graphs 
of NCQFT may be interpreted either as ``thicker particle world-lines" or as ``simplified open strings world-sheets" in which only the ends of strings appear but not yet their internal
oscillations. 

\subsection{Scalar models on the Moyal space}

The noncommutative Moyal space is defined in even dimension $d$ by
\beqa
[x^\mu, x^\nu]_\star=\imath \Theta^{\mu \nu},
\eeqa
where $\Theta$ 
is an antisymmetric $d/2$ by $d/2$ block-diagonal matrix with blocks:
\beqa
\label{theta}
\begin{pmatrix}
   0 &\theta \\   
   -\theta & 0 
  \end{pmatrix}  
\eeqa
and we have denoted by  $\star$ the Moyal-Weyl product
\beqa
\label{moyal-product} 
 (f\star g)(x)=\int \frac{d^{4}k}{(2\pi)^{4}}d^{4}y\, f(x+{\textstyle\frac 12}\Theta\cdot
  k)g(x+y)e^{\imath k\cdot y} .
\eeqa
Note that in the limit $\theta\to 0$ this product becomes the ordinary commutative product of functions.

\subsubsection{The ``naive'' model}
\label{naive}

The simplest field theory on this space consists
in replacing the ordinary commutative local product of fields   
 by the Moyal-Weyl product
\beqa
\label{act-normala}
S[\phi]=\int d^d x (\frac 12 \partial_\mu \phi \star \partial^\mu \phi +\frac
12 \mu^2 \phi\star \phi  + \frac{\lambda}{4} \phi \star \phi \star \phi \star \phi).
\eeqa 

In momentum space the action (\ref{act-normala}) writes
\beqa
\label{act-normala-p}
S[\phi]=\int d^d p (\frac 12 p_\mu \phi  p^\mu \phi +\frac
12 \mu^2 \phi  \phi  + V (\phi,\theta)).
\eeqa
where $V(\phi,\theta)$ is the corresponding potential.

An important consequence of the use of the non-local product $\star$ is that the interaction part
no longer preserves the invariance under permutation of external fields. This invariance is 
restricted to cyclic permutations. Furthermore, there exists a basis - the matrix base - of the Moyal algebra where the Moyal-Weyl product takes the form of an ordinary (infinite) matrix product. For these reasons
the associated Feynman graphs are ribbon graphs, that is propagators should be drawn as ribbons.



\begin{figure}
\begin{center}
\includegraphics[scale=0.9]{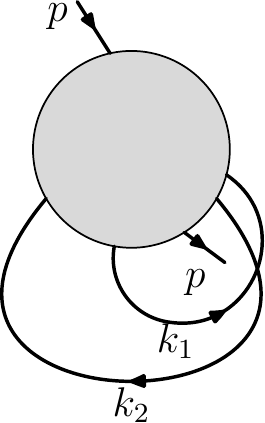}
\end{center}
\caption{An example of a rosette with two flags.
The crossings of edges $k_1$ and $k_2$ indicate the non trivial genus (here $g=1$).}\label{roz}
\end{figure}

In \cite{filk} several contractions on such a Feynman 
graph were defined. In particular the ``first Filk move" 
is the contraction introduced in subsection \ref{delcontractribbon}.
Repeating this operation for the $V-1$ edges of a spanning tree, one obtains a {\it rosette} (see Figure \ref{roz}).

Note that the number of faces or the genus of the graph does not change 
under contraction. There is no crossing between edges for a planar rosette. The example of Figure \ref{roz} corresponds thus to a non-planar graph (one has crossings between the edges $k_1$ and $k_2$). This pair is called a {\it nice crossing} pair. 

The notions expressed in the previous section (namely the Green and Schwinger functions or the perturbation theory concepts) remain the same as in QFT. Usual Feynman graphs are simply replaced
by ribbon Feynman graphs.

Recall that this ``naive model'' \eqref{act-normala} is not renormalizable
in $d=4$. This is due to a new type of non-local divergence at the level of the $2-$point 
function - the UV/IR mixing \cite{MinSei}.

\subsubsection{A translation-invariant renormalizable scalar model}
\label{GMRT}

In order to restore renormalizability at $d=4$, the propagator can be modified in the following way 
\cite{noi}
\beqa
\label{revolutie}
S_{GMRT}[\phi]=\int d^d p \; (\frac 12 p_{\mu} \phi  p^\mu \phi  +\frac
12 m^2  \phi  \phi   
+ \frac 12 a  \frac{1}{\theta^2 p^2} \phi  \phi  
+ \frac{\lambda }{4} \phi  \star \phi  \star \phi  \star \phi ),
\eeqa
where $a$ is some dimensionless parameter which is taken in the interval $0<a\le \frac 14 \theta^2 m^4$.

The corresponding propagator writes in momentum space
\beqa
\label{propa-rev}
C_{GMRT}=\frac{1}{p^2+\mu^2+\frac{a}{\theta^2 p^2}} \, .
\eeqa
In \cite{noi}, this model was proved to be renormalizable at any order in perturbation theory. Furthermore, its renormalization group flows \cite{beta-GMRT} 
were calculated; a mechanism for taking the commutative limit has been proposed \cite{limita} (for a review on all these developments, see \cite{review-io}).

\subsection{The NC Parametric representation}

In this subsection we present the implementation of the parametric representations for the noncommutative scalar models introduced in the previous subsection.

To keep track of the cyclic ordering at the vertex it is convenient
to detail the incidence matrix $\e_{ev}$ into a more precise 
incidence tensor $\e^v_{ei}$ where $i = 1,...,4$ indexes the four
corners of the Moyal vertex. As before it is 1 if the edge $e$ starts
at corner $i$ of vertex $v$, -1 if it exits at that corner, and 0 otherwise.

To implement the parametric representation we follow subsection \ref{paramQFT}. The propagator
remains the same as in QFT, but the contribution of a vertex $v$ now corresponds to a Moyal kernel.
In momentum space it writes using again summation over repeated indices
\beqa
\label{v1}
\delta (\sum_{i=1}^4 \e^v_{ei}k_e )e^{- \frac i2\sum_{1\le
    i <j\le 4}\e^v_{ei}k_e\Theta \e^v_{ej}k_e} .
\eeqa
\noi
By $k_i\Theta k_j$ we denote $k_i^\mu \Theta_{\mu\nu} k^\nu_j$.
The $\delta-$function appearing in the vertex contribution \eqref{v1} is
nothing but the usual momentum conservation. It can be 
written as an integral over a new variable $\tilde x_v$, called {\it hyperposition}. One associates such a variable to any Moyal vertex, even though this vertex is non-local: 
\beqa
\delta(\sum_{i=1}^4 \e^v_{ei}k_e ) = \int  \frac{d \tilde x_v'}{(2 \pi)^4}
e^{i\tilde x_v'(\sum_{i=1}^4 \e^v_{ei}k_e )}
=\int  \frac{d \tilde x_v}{(2 \pi)^4}
e^{\tilde x_v\sigma(\sum_{i=1}^4 \e^v_{ei}k_e )}.\label{pbar1}
\eeqa
where $\sigma$ is a $d/2$ by $d/2$ block-diagonal matrix with blocks:
\beqa
\label{sigma}
\sigma = 
\begin{pmatrix}
0 & -i  \\
i & 0 
\end{pmatrix}.
\eeqa
Note that to pass from the first to the second line in \eqref{pbar1}, the change of variables $i \tilde x_v'=\tilde x_v \sigma$ has Jacobian 1.


\subsubsection{The ``naive'' model}

Putting now together the contributions of all the internal momenta and vertices, one has the following parametric representation:
\beqa
\label{param-naiv}
&&{\cal A}_G^T(p_1,\ldots,p_N)=K_G^T\int \prod_{e,e'=1}^E d^d k_e d\alpha_e 
e^{-\alpha_e (k_e^2+m^2)}\\
&&\prod_{v=1}^{V-1}\int d^d \tilde x_v 
e^{i \tilde x_v (\sum_{i=1}^4 \e^v_{ei}k_e)}
e^{-\frac i2 \sum_{i <j} \e_{ei}^v k_e \Theta \e_{e'j}^v k_{e'}}
\eeqa
where we have denoted by
 $K_G^T$ some inessential normalization constant. Furhermore note that in the integrand above we have denoted, to simplify the notations, by $k_e$ or $k_{e'}$ momenta which can be both internal or external.

\subsubsection{The translation-invariant model}

The parametric representation of the model \eqref{revolutie} was analyzed in \cite{param-GMRT}. 
This representation is intimately connected to the one of the model \eqref{act-normala}
(see the previous subsubsection) for the following reason. One can rewrite the 
propagator \eqref{propa-rev} as 
\beqa
\frac{1}{A+B}=\frac 1A - \frac 1A B \frac{1}{A+B}
\eeqa
for 
\beqa
\label{AB}
A=p^2+m^2,\ \ B=\frac{a}{\theta^2 p^2}.
\eeqa
Thus, the propagator \eqref{propa-rev} writes
\beqa
\label{propa2}
C_{GMRT}&=&\frac{1}{p^2+m^2}-\frac{1}{p^2+m^2}\frac{a}{\theta^2 p^2 (p^2+m^2)+a},\nonumber\\
&=&
\frac{1}{p^2+m^2}-\frac{1}{p^2+m^2}\frac{a}{\theta^2 (p^2 +m_1^2)(p^2+m^2_2)}
\eeqa
where $-m_1^2$ and $-m_2^2$ are the roots of the denominator of the second term in the LHS (considered as a second order equation in $p^2$, namely
$\frac{-\theta^2 m^2\pm \sqrt{\theta^4 m^4 - 4 \theta^2 a}}{2\theta^2}<0$. 
Note that the form \eqref{propa2} allows us already to write an integral representation of the propagator $C(p,m,\theta)$. Nevertheless, for the second term one would need a triple integration over some set of Schwinger parameters:
\beqa
\label{param}
C_{GMRT}&=&\int_0^\infty d\alpha e^{-\alpha (p^2+m^2)},\\
&-& \frac{a}{\theta^2} \int_0^\infty \int_0^\infty d\alpha d\alpha^{(1)} d\alpha^{(2)} e^{-(\alpha+\alpha^{(1)}+\alpha^{(2)})p^2} e^{-\alpha m^2} e^{-\alpha^{(1)} m_1^2}e^{-\alpha^{(2)} m_2^2}.\nonumber
\eeqa
Instead of that one can use the following formula:
\beqa
\frac{1}{p^2+m_1^2}\frac{1}{p^2+m_2^2}=\frac{1}{m_2^2-m_1^2}(\frac{1}{p^2+m_1^2}-\frac{1}{p^2+m_2^2}).
\eeqa
This allows to write the propagator \eqref{propa2} as
\beqa
C_{GMRT}=\frac{1}{p^2+m^2}-\frac{a}{\theta^2 (m_2^2-m_1^2)}\frac{1}{p^2+m^2}(\frac{1}{p^2+m_1^2}-\frac{1}{p^2+m_2^2}).
\eeqa
This form finally allows to write down the following integral representation:
\beqa
\label{propa-int2}
C_{GMRT}&=&\int_0^\infty d\alpha e^{-\alpha (p^2+m^2)}
- \frac{a}{\theta^2 (m_2^2-m_1^2)} \int_0^\infty \int_0^\infty d\alpha d\alpha_1 
e^{-(\alpha+\alpha_1)p^2-\alpha m^2 -\alpha_1 m_1^2}\nonumber \\
&+& \frac{a}{\theta^2 (m_2^2-m_1^2)} \int_0^\infty \int_0^\infty d\alpha d\alpha_2 e^{-(\alpha+\alpha_2)p^2} e^{-\alpha m^2} e^{-\alpha_2 m_2^2}. 
\eeqa

Let us also remark that the noncommutative propagator $C_{GMRT}$ is bounded by the ``usual'' commutative propagator $C(p,m)$
\beqa
\label{limita}
C_{GMRT} \le C(p,m).
\eeqa

\medskip

Using now \eqref{param}, the parametric representation of the model \eqref{revolutie} is thus a sum of $2^E$ terms coming from the development of the $E$ internal propagators. 
Each of these terms has the same form of the one of polynomials in the previous subsection. The only differences comes from 
\begin{itemize}
\item the proper substitution of the set of Schwinger $\alpha$ parameters
\item  the mass part.
\end{itemize}
One has
\beqa
\label{plr}
&&{\cal A}_G^T=K_G^T \left( \int \prod_{i=1}^L d \alpha_i \frac{1}{[U(\alpha)]^{\frac D2}} e^{\frac{-V(\alpha, p)}{U(\alpha)}} e^{-\sum_{i=1}^L \alpha_i m^2 }\right.\\
&& + (-\frac{a}{\theta^2})^{L-1}\sum_{j_1=1}^L \int d\alpha_{j_1} \prod_{i\ne j_1,\, i=1}^L d\alpha_i d\alpha_i^{(1)} d\alpha_i^{(2)} \frac{1}{[U(\alpha_i+\alpha_i^{(1)}+\alpha_i^{(2)}, \alpha_{j_1})]^\frac d2} \nonumber\\
&& \ \ \ e^{-\frac{V(\alpha_i+\alpha_i^{(1)}+\alpha_i^{(2)}, \alpha_{j_1},p)}{U(\alpha_i+\alpha_i^{(1)}+\alpha_i^{(2)}, \alpha_{j_1})}} 
e^{-\sum_{i=1}^L \alpha_i m^2 } e^{-\sum_{i\ne j_1,\, i=1}^L \alpha_i^{(1)} m_1^2 }e^{-\sum_{i\ne j_1,\, i=1}^L \alpha_i^{(2)} m_2^2 }\nonumber\\
&& + (-\frac{a}{\theta^2})^{L-2}\sum_{j_1<j_2,\, j_1,j_2=1}^L \int d\alpha_{j_1}d\alpha_{j_2} \prod_{i\ne j_1,j_2,\, i=1}^L d\alpha_i d\alpha_i^{(1)} d\alpha_i^{(2)} \nonumber\\
&& \ \ \ \ \ \ \ \ \ \ \ \ \ \ \ \ \ \ \ \ \ \ \ \ \ \ \  \frac{1}{[U(\alpha_i+\alpha_i^{(1)}+\alpha_i^{(2)}, \alpha_{j_1}, \alpha_{j_2})]^\frac d2} \nonumber\\
&& \ \ \ \ \ \ \ \ \ \ e^{-\frac{V(\alpha_i+\alpha_i^{(1)}+\alpha_i^{(2)}, \alpha_{j_1},\alpha_{j_2}, p)}{U(\alpha_i+\alpha_i^{(1)}+\alpha_i^{(2)}, \alpha_{j_1}, \alpha_{j_2})}}
e^{-\sum_{i=1}^L \alpha_i m^2 } e^{-\sum_{i\ne j_1j_2,\, i=1}^L \alpha_i^{(1)} m_1^2 }e^{-\sum_{i\ne j_1,j_2\, i=1}^L \alpha_i^{(2)} m_2^2 }
\nonumber\\
&& + \ldots +\nonumber\\
&&  + (-\frac{a}{\theta^2})^{L}\int  \prod_{i=1}^L d\alpha_i d\alpha_i^{(1)} d\alpha_i^{(2)} \frac{1}{[U(\alpha_i+\alpha_i^{(1)}+\alpha_i^{(2)})]^\frac d2} e^{-\frac{V(\alpha_i+\alpha_i^{(1)}+\alpha_i^{(2)},p)}{U(\alpha_i+\alpha_i^{(1)}+\alpha_i^{(2)})}}\nonumber\\
&& \left. \ \ \ \ \ \ \ \ \ \ \ \ \ \ \ \ \ \ \ \ \ \ \ \ \ \ \ 
e^{-\sum_{i=1}^L \alpha_i m^2 }e^{-\sum_{i=1}^L \alpha_i^{(1)} m_1^2 }e^{-\sum_{i=1}^L \alpha_i^{(2)} m_2^2 }
\right).\nonumber
\eeqa

\subsection{Deletion/contraction for the NC Symanzik polynomials}

In this subsection we give some results relating the Bollob\'as-Riordan polynomial and the parametric representations of the noncommutative scalar models introduced here.

\subsubsection{The ``naive'' model}

As in the commutative case, we have to perform a Gau\ss ian integral in a $d(E+V-1)$ dimensional space. 
Consider a ribbon graph $G$ with a root $\bar v$.

We introduce the condensed notations analog to \eqref{mainformb}-\eqref{mainformq}
\bea A_G (p) = \int   \prod_e
d\alpha_e    e^{-\alpha_e m^2} \int d^d \tilde x d^d p 
e^{-  Y X Y^t} 
\eea 
where 
\bea \label{defMPQ1NC} 
Y = \begin{pmatrix}
k_e & \tilde x_v & p_e & \tilde x_{\bar v} \\
\end{pmatrix} 
\ \ , \ \   X= \begin{pmatrix} Q & -iR^t \\ -iR & M \\
\end{pmatrix}\ .
\eea
$Q$ is an  $d(E + V-1)$-dimensional square matrix. We have denoted by $p_e$ the external momenta and by $\tilde x_{\bar v}$ the hyperposition associated to the root vertex $\bar v$. The matrix $R$ is a $d(N+1)\times d(E+V-1)$ dimensional matrix and $M$ is a $d(N+1)$ dimensional square matrix
representing the Moyal couplings between the external momenta and the root vertex.

Gau\ss ian integration gives, up to inessential constants:
\bea\label{defMPQ2NC} A_G (p) = \int \prod_e d\alpha_e e^{-\alpha_e m^2}
\frac{1}{\det Q^{d/2}}    e^{ - P R Q^{-1}  R^{t}   P^t  }
\eea
where $P$ is a line matrix regrouping the external momenta (and the hyperposition associated to the root vertex).

The determinant of the matrix $Q$
defines therefore the first Symanzik NC-polynomial $U^\star$ and the product of the
matrices $R$ and inverse of $Q$ defines the quotient of the second
Symanzik polynomial $V^\star $ by $U^\star$ where the star recalls the Moyal product
used to define this NCQFT.

Let us  calculate first the determinant of $Q$. 
One has
\beqa
\label{important}
Q= D\otimes 1_d + A \otimes \sigma
\eeqa
where $D$ is a diagonal matrix with 
coefficients $D_{ee} =\alpha_e$, for $e=1,\ldots, E$ and $D_{vv} = 0$ for the rest, $v=1,\ldots, V-1$.  
$A$ is an antisymmetric matrix. In \cite{GurauRiv} it was noted that, for such a matrix
\beqa
\det Q = \det (D+A)^d.
\eeqa 
This implies, as in the commutative case, that
\beqa
U^\star=\det (D+A).
\eeqa
Factoring out powers of $i$ one has
\begin{equation}
\label{Qnaiv}
\det (D+A) = \det 
\begin{pmatrix}
\alpha_1 & f_{12}& & & -{\sum_{i=1}^4\epsilon^v_{e i}}\\
 -f_{12}& \alpha_2 & \\& & \ldots&\\ & & & \ldots& \\
{\sum_{i=1}^4\epsilon_{e i}^v} & & & & 0\\
\end{pmatrix}.
\end{equation}
The difference with the commutative case comes from the 
non-trivial antisymmetric coupling between the $E$ edges variables. 
It corresponds to an $E$ dimensional square matrix $F$ with matrix elements
\beqa
\label{F}
f_{ee'}=-\frac{\theta}{2} \sum_{v=1}^{n} \sum_{i,j=1}^4 \omega(i,j)\e_{ei}^v \e_{e'j}^v, \ \forall e<e',\ e,e'=1,\ldots, E
\eeqa
where $\omega$ is an antisymmetric matrix such that $\omega (i,j)=1$ if $i<j$. This matrix takes into account the antisymmetric character of $\Theta$ in $k_\mu \Theta^{\mu \nu} p_\nu$.

Using again Lemma \ref{quasipfaff}
\begin{eqnarray}
\det (D+A)&=&  \int \prod_{i,e} d\omega_i d\chi_i d\omega_e d\chi_e 
\nonumber\\&&e^{- \sum_e \alpha_e\chi_e
\omega_e} e^{-\sum_{e,v}\chi_e
\epsilon_{ev}\chi_v   + \chi \leftrightarrow  \omega  }
 e^{-\sum_{e,e'}\chi_e
f_{ee'}\chi_{e'}   + \chi \leftrightarrow  \omega  }.
\end{eqnarray}
Note that the last term above represents the difference with the commutative case.

We have the exact analog of Theorem \ref{grasstheo}
to prove a deletion-contraction rule.
\begin{theorem}\label{grasstheo2}
For any semi-regular edge $e$
\bea \label{delcontr3}
 \det (D+A)_G= \alpha_e \det (D+A)_{G- e} + \det (D+A)_{G/ e}.
\eea
\end{theorem}

\prf  
We pick up a semi-regular edge $e$ entering $v_1$ and exiting $v_2$. 
Thus it exists some $i$ and $j$  with $\epsilon^{v_1}_{ei} = +1, \epsilon^{v_2}_{ej} = -1$. We expand
\be e^{- \alpha_e\chi_e \omega_e} = 1 +\alpha_e\omega_e\chi_e. 
\ee
leading to two contributions, which we denote respectively by $\det Q_{G,e,1}$ and 
$\det Q_{G,e,2}$.
For the first term, since one must saturate the $\chi_e$ 
and $\omega_e$ integrations, one has to keep the $\chi_e (\chi_{v_1} - \chi_{v_2} + \sum_{\tilde e} f_{e\tilde e}\chi_{\tilde e}) $ term and the similar $\omega$ term. 
Note that the sum is done on all the edges $\tilde e$ hooking to any of the vertices $v_1$ and $v_2$ and with whom the edge $e$ has no trivial Moyal oscillation factor.
One has
\begin{eqnarray}
 \det Q_{G,e,1}&=& \int  \prod_{e'\ne e,v}  d \chi_{e'} d\chi_v  d\omega_{e'} d\omega_v  \nonumber\\
&& 
 (\chi_{v_1} - \chi_{v_2}+ \sum_{\tilde e} f_{e\tilde e}\chi_{\tilde e})   
(\omega_{v_1} - \omega_{v_2}+ \sum_{\tilde e} f_{e\tilde e}\omega_{\tilde e})\nonumber\\
&&
e^{-\sum_{e'\ne  e }\alpha_e'\chi_{e'}
\omega_{e'}} e^{-\frac{1}{4}\sum_{e' \ne e ,v}   \chi_{e'}
\epsilon_{e'v}\chi_v+ \chi \leftrightarrow \omega   }.
\end{eqnarray}
As in the commutative case, we now perform the trivial triangular  change of variables  with unit Jacobian:
\be  \hat \chi_{v_1} = \chi_{v_1} - \chi_{v_2}+ \sum_{\tilde e} f_{e\tilde e}\chi_{\tilde e}, \ \   \hat \chi_{v} = \chi_{v}\ \   \emph{for } \ v \ne v_1,
\ee
and the same change for the $\omega$ variables. What happens now is analogous to the commutative case, with the difference that the last term in the definition of $\hat \chi_{v_1}$ will lead to the reconstruction of  the Moyal oscillation factors of the edges hooking to $v_1$ with the edges hooking to $v_2$. This completes the ribbon contraction, thus 
$ \det Q_{G,e,1} =  \det Q_{G/e}$. The second term
$\det Q_{G,e,2}$ with the $ \alpha_e\omega_e\chi_e$
factor is even easier. We must simply put to 0 all terms involving the $e$ label, hence trivially
$ \det Q_{G,e,2} = \alpha_e \det Q_{G-e}$. 
\qed

We need now to compute $U^\star$ on terminal forms after contracting/deleting all
semi-regular edges, that is compute $U^\star$ on  a rosette graph ${\cal R}$.
This is done by using the double contraction introduced in the previous
section.

Consider a nice crossing of ${\cal R}$ between two edges $e_1$ and $e_2$
with parameters $\alpha_1$ and $\alpha_2$. It leads to a contribution
\beqa\label{terminal}
U_{{\cal R}}^\star = ( \alpha_1\alpha_2+\frac 14 \theta^2 ) U_{{\cal R}/e_1e_2}
\eeqa
where we recall that the contracted rosette ${\cal R}/e_1e_2$
is obtained by deleting $e_1$ and $e_2$ from ${\cal R}$ and interchanging 
the half-edges encompassed by $e_1$ with the ones encompassed by $e_2$, 
see Figure \ref{filk3}. The procedure 
can be iterated on ${\cal R}/e_1e_2$ until after $g({\cal R})$ 
double contractions a planar rosette with $2E({\cal R}) -2g({\cal R})$
is reached, for which  $F=0$ and for which the terminal form is 
$\prod_e \alpha_e$ as in the commutative case. 

Remark that the main difference with the commutative case is the inclusion 
of the $\theta^2$ term in the terminal form evaluation 
\eqref{terminal}. 
This type of \emph{genus-term} has no analog in the commutative case.

\begin{example}
Consider the graph of Figure \ref{graf-NP}.
\begin{figure}
\begin{center}
\includegraphics[scale=0.9]{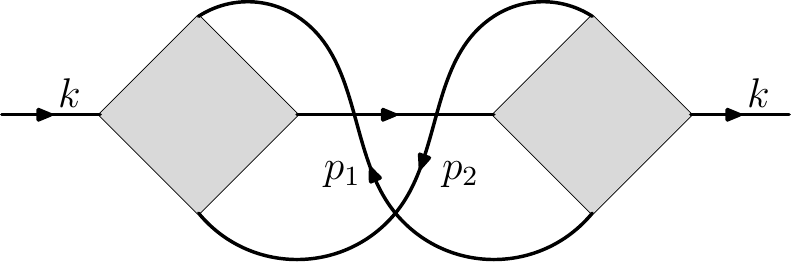}
\caption{An example of a non-planar graph, $g=1$.}\label{graf-NP}
\end{center}
\end{figure}
Its first Symanzik polynomial is \cite{param-GMRT}
\beqa
\label{tot}
\alpha_1\alpha_2+\alpha_1\alpha_3+\alpha_2\alpha_3+\frac 14 \theta^2.
\eeqa
Choosing $\alpha_3$ as a regular edge leads to a contracted graph 
where the pair of edges $\alpha_1$ and $\alpha_2$ realizes a nice crossing. We thus have a contribution to the first polynomial
\beqa
\label{p1}
\alpha_1\alpha_2 +\frac 14 \theta^2 .
\eeqa
The deleted part then follows as in the commutative case leading to a contribution
\beqa
\label{p2}
\alpha_3\alpha_1+\alpha_3\alpha_2.
\eeqa
Putting together \eqref{p1} and \eqref{p2} leads to the expected result \eqref{tot}.
\end{example}

\medskip

Let us now give the following definition:

\begin{definition}
A $\star$-tree of a connected graph $G$ is a subset of edges with  one boundary.
\end{definition}

This definition allows to write a $\star$-tree in some graph of genus $g$ as 
an ordinary tree plus at most $g$ pairs of ``genus edges'' (where by ``genus edges'' we 
understand pairs of edges which make a recursive succession of nice crossings under
double contractions on the rosette obtained after contracting the edges of the tree in the graph).

\begin{example}
For the graph of Figure \ref{graf-NP}, the $\star$-trees are the ordinary trees $\{1\}$, $\{2\}$, $\{3\}$ 
and the tree plus one pair of genus edges, namely $\{1,2,3\}$ which is the whole graph.
\end{example}

In \cite{MinSei}, the following general expression for the first 
polynomial $U$ of the ``naive'' noncommutative model was given
\beqa
\label{min1}
U^\star(\alpha_1,\ldots,\alpha_E)=\left(\frac{\theta}{2}\right)^{b} \sum_{{\cal T}^\star\; \star-{\mathrm tree}} \prod_{e\notin {\cal T}^\star} 2 \frac{\alpha_e}{\theta},
\eeqa
where we have denoted by 
\beqa
\label{b}
b=F-1+2g
\eeqa 
the number of loops of $G$. Note that the factor $2$ above is the one which matches our conventions.

Let us now give a proof of the formula \eqref{min1}. 
Consider the following lemma:

\begin{lemma}
{\bf (Lemma III.2 of \cite{GurauRiv})}
Let $D=(d_i\delta_{i j})_{i,j\in\{1,\dotsc,D\}}$ 
be diagonal and $A=(a_{i j})_{i,j\in\{1,\dotsc,D\}}$ 
be such that $a_{ii}=0$. Then
\beqa      
\det(D+A)=\sum_{K\subset \{1,\dotsc ,N\}}\det(B_{\hat{K}}) \prod_{i\in K}a_i 
\eeqa
where $A_{\hat{K}}$ is the matrix obtained from $A$ by deleting the lines and 
columns with indices in $K$.
\end{lemma}

The particular form \eqref{important} of the matrix $Q$ allows thus to use this Lemma to calculate its determinant ({\it i.e.} the polynomial $U$). Factoring out $\frac {\theta}{2}$ on the first $E$ lines and then $\frac{2}{\theta}$ on the last $V-1$ columns, one has
\beqa
U^\star (\alpha)=\left(\frac{\theta}{2}\right)^b\sum_{K\subset\{1,\ldots,E\}} \det A_{\hat K} \prod_{e\in K} 2 \frac{\alpha_e}{\theta}
\eeqa
where we have used that
$$ b - E = - (V-1).$$
Note that the set $K$ on which one sums up corresponds to a set of edges of the graph; this comes from the fact that the last $V-1$ entries on the diagonal of the matrix $A$ are equal to $0$. 
In \cite{GurauRiv} (see Lemma III.4) it is proven, using a non trivial triangular change of Grassmanian variables that a determinant of type $A_{\hat K}$ is not vanishing if and only if it corresponds to a graph with only one face. This means that the complement of the subset of edges $K$ must be a $\star-$tree, $\bar K={\cal T}^*$. Furthermore, one has
\beqa
\prod_{\bar {\cal T}^\star} \alpha_e = \prod_{e\notin {\cal T}^\star} \alpha_e.
\eeqa

\subsubsection{The translation-invariant model}

The relation of the parametric representation of the model \eqref{revolutie} to the Bollob\'as-Riordan polynomials follows the one of the ``naive'' model \eqref{act-normala} presented above. This is an immediate consequence of the intimate relationship between the parametric representation of these two noncommutative models, a relationship explained in the previous subsection.

\subsection{The second polynomial for  NCQFT}

In this section we prove the form of the second polynomial for the model 
\eqref{act-normala} (both its real and imaginary part, as we will see in the sequel). 
We then relate this second polynomial to the Bollob\'as-Riordan polynomial.

From 
\eqref{defMPQ2NC} it follows directly that
\beqa
\frac{V^\star(\alpha,p)}{U^\star(\alpha)}=-P 
R Q^{-1}  R^{t} 
P^t  
\eeqa
where we have left aside the matrix $M$ coupling the external momenta to themselves. Note that the matrix $R$ couples the external momenta (and the hyperposition associated to the root vertex) to the internal momenta and the remaining $V-1$ hyperpositions. This coupling is done  in an analogous way to
 the coupling of the internal momenta with the respective variables.

We can thus state that the $V$ polynomial is given, as in the commutative case, by the inverse $Q^{-1}$ of the matrix $Q$ giving the $U$ polynomial.
The particular form \eqref{important} of the matrix $Q$ leads to 
\bea
Q^{-1}&=&\frac 12 \left((D+A)^{-1}+(D-A)^{-1}\right)\otimes 1_d \nonumber\\
&+& \frac 12 \left((D+A)^{-1}-(D-A)^{-1}\right)\otimes \sigma
\eea
Thus, the polynomial $V$ has both a real $V^R$ and an imaginary part $V^I$.

In the case of the commutative theories, the imaginary part above disappears. This is a consequence of the fact that the matrix $F$, coupling through the Moyal oscillations the internal momenta, vanishes for $\theta=0$.

Let the following definitions.

\begin{definition}
A two $\star$-tree is a subset of edges with two boundaries.
\end{definition}

Furthermore, let $K$ a subset of lines of the antisymmetric matrix $A$.
Let Pf$(A_{\hat K\hat\tau})$ be the Pfaffian of the antisymmetric matrix  obtained from
$A$  by deleting the edges in the set $K\cup\{\tau\}$ for $\tau\notin K$.  We also define $\varepsilon_{K,\tau}$ to be the signature of the permutation obtained from $(1,\ldots, E)$ by extracting the positions belonging to $K\cup\{\tau\}$ and replacing them at the end in the order:
$$ 1,\ldots, E\to 1,\ldots, \hat i_1,\ldots, \hat i_p ,\ldots, \hat i_\tau,\ldots, E, i_\tau, i_p,\ldots , i_1.$$

We now prove a general form for both the real and the imaginary part of the polynomial $V^\star$,
noted ${\cal X}^{\star}$ and ${\cal Y}^{\star}$.

\begin{theorem}
\label{thr}
The real part of the second Symanzik polynomial writes
\beqa
\label{vr}
{\cal X}^{\star}=\left(\frac{\theta}{2}\right)^{b+1}\sum_{{\cal T}^\star_2\; 2-\star \;{\mathrm{tree}}}\prod_{e\notin {\cal T}_2^\star}2\frac{\alpha_e}{\theta}(p_{{\cal T}^\star_2})^2,
\eeqa
where $p_{{\cal T}^\star_2}$ is the sum of the momenta entering one of the two faces of the $2$ $\star-$tree ${\cal T}_2^\star$.
\end{theorem}
Note that by momentum conservation, the choice of the face in the above theorem is irrelevant. Furthermore, let us emphasize on the fact that, being on the submanifold $p_G=0$, an equivalent writing of \eqref{vr} is
\beqa
{\cal X}^{\star}=-\frac12 \left(\frac{\theta}{2}\right)^{b+1}\sum_{{\cal T}^\star_2\; 2-\star \;{\mathrm{tree}}}\prod_{e\notin {\cal T}_2^\star}2\frac{\alpha_e}{\theta}p_v\cdot p_{v'},
\eeqa
where $p_v$ (and resp. $p_{v'}$) is the total momenta entering one of the two faces of the $2$-$\star$ tree.

\medskip

\noindent
{\it Proof.} We base our proof on the following lemma:

\begin{lemma} ({\bf Lemma IV.1 of \cite{GurauRiv}}) \\
The real part of the polynomial $V^\star$ writes
\beqa
\label{lemar}
{\cal X}^{\star}=\sum_{K}\prod_{i\notin K} d_i \left(\sum_{e_1}p_{e_1}\sum_{\tau \notin K} R_{e_1 \tau} \varepsilon_{K\tau}{\mathrm {Pf}}(A_{\hat K\hat \tau})\right)^2
\eeqa
where $d_i$ are the elements on the diagonal of the matrix $Q$. Furthermore, when $|K|\in\{E-1,E\}$ the matrix with deleted lines is taken to be the empty matrix, with unit Pfaffian.
\end{lemma}
Note that, as before, since the matrix $Q$ has vanishing entries on the diagonal for the last $V-1$ entries  the subsets $K$ are nothing but subsets of edges. The empty matrix obtained from deleting all the first $E$ edges in the graph corresponds to the graph with no internal edges but only disconnected vertices. Each of these disconnected components has one boundary; hence the Pfaffian is non-vanishing.

Note that the Pfaffian in \eqref{lemar} disappears iff the corresponding graph has $1$ boundary (see above). This means that
$ K\cup \{\tau\}$
is the complement of a $\star-$tree ${\cal T}^\star$:
\beqa
{\overline{K\cup \{\tau\}}}={\cal T}^\star.
\eeqa
 Hence the subset $K$ is the complement of a $\star$-tree plus an edge (just like in the commutative case). Adding an extra edge to a $\star$-tree represents an increase of the number of boundaries by one unit. Hence, the subset of edges $K$ above is the complement of some $2$ $\star-$tree ${\cal T}_2^\star$
\beqa
\bar K= {\cal T}_2^\star.
\eeqa
As before, one has
$$ \prod_{e\in K} \alpha_e=\prod_{e\notin {\cal T}_2^\star} \alpha_e.$$
The diagonal terms in the matrix $Q$ are again the parameters $\alpha_e$. Factoring out $\frac{\theta}{2}$ factors on the lines of the matrices corresponding to the edges of the graph and then $\frac{2}{\theta}$ for the lines of the matrices corresponding to the vertices. The extra factor $\theta/2$ corresponds to the extra edge $\tau$.

\medskip

Let us now investigate the square root of the momenta combination entering \eqref{lemar}. Note that the matrix element $R_{e_1\tau}$ is not vanishing only for external momentum $p_{e_1}$ which has a Moyal oscillation with the internal momenta associated to the edge $\tau$. It is this edge $\tau$ which actually creates the extra boundary. Thus the sum on the external momenta in \eqref{lemar} is nothing but the sum of the momenta entering one of the two boundaries. By a direct verification, one can explicitly check the signs of the respective momenta in \eqref{lemar}, which concludes the proof. \qed

\begin{example}
For the graph of Figure \ref{graf-NP}, the second polynomial is
\beqa
V^{\star}(\alpha,p)=\alpha_1 \alpha_2 \alpha_3 p^2 + \frac 14 (\alpha_1+\alpha_2+\alpha_3)\theta^2p^2.
\eeqa
\end{example}

\bigskip

Let us now investigate the form of the imaginary part ${\cal Y}^{\star}$. One has the following theorem:

\begin{theorem}
\label{thi}
The imaginary part of the second Symanzik polynomial writes
\beqa
\label{vim}
{\cal Y}^{\star}(\alpha,p)=\left(\frac{\theta}{2}\right)^{b}\sum_{{\cal T}^\star\; \star \;{\mathrm{tree}}}\prod_{e\notin {\cal T}^\star}2\frac{\alpha_e}{\theta} \psi(p),
\eeqa
where $\psi(p)$ is the phase obtained by following the momenta entering the face of the $\star-$tree ${\cal T}^\star$ 
as if it was a Moyal vertex.
\end{theorem}
{\it Proof.} The proof follows closely the one of Theorem \ref{thr}. Nevertheless, the equivalent of \eqref{lemar} is now (see again \cite{GurauRiv})
\beqa
{\cal Y}^{\star}(\alpha,p)&=&\sum_{K}\prod_{e\notin K} d_i \epsilon_K {\mathrm{Pf}}(A_{\hat K}) \left(\sum_{e_1,e_2}\left(\sum_{\tau,\tau'}R_{e_1\tau}\epsilon_{K\tau\tau'}{\mathrm{Pf}}(A_{\hat K \hat \tau \hat \tau'})R_{e_2\tau'}\right)p_{e_1}\sigma p_{e_2}\right) \nonumber
\\
&&\eeqa
where $d_i$ are the elements on the diagonal of the matrix $Q$. Since we look this time for sets such that ${\mathrm{Pf}}(B_{\hat K})$ is non-vanishing, this implies as above that $K$ is the complement of some $\star$-tree ${\cal T}^\star$. 
Furthermore one needs to consider the two extra edges $\tau$ and $\tau'$. It is possible from the initial $\star-$tree above to erase these two more edges such that the Pfaffian ${\mathrm Pf}B_{\hat K \hat \tau \hat \tau'}$ is non-vanishing. Indeed, if the $\star$-tree is a tree, by erasing two more edges of it we obtain a graph with $3$ disconnected components, each of it with a single boundary; the corresponding Pfaffian will be non-vanishing. Summing up on all these possibilities leads to the Moyal oscillations of the external momenta (the one which disappears when truncating the graph). If the $\star-$tree is formed by a tree and some pair of genus edges we can always delete further the pair of genus edges and remain with the regular tree. Obviously the corresponding Pfaffian is again non-vanishing (since it corresponds to a graph with only one boundary).
\qed

Note that the form of the real part and of the imaginary one of the polynomial $V^\star$ are qualitatively different. Indeed, the real part contains some square of a sum of incoming external momenta, while the imaginary one contains a phase involving the external momenta.

\bigskip

Let us end this section by stating that the second noncommutative Symanzik polynomial 
also obeys the deletion-contraction rule. The proof is exactly like in the commutative case,
a straightforward rereading of Theorem \ref{grasstheo2}.

\subsection{Relation to multivariate Bollob\'as-Riordan polynomials}

In the previous subsections, we have identified the first Symanzik polynomial of a connected graph in a scalar NCQFT as the first order in $w$ of the multivariate Bollob\'as-Riordan polynomial,
\begin{equation}
U^{\star}_{G}(\alpha,\theta)=(\theta/2)^{E-V+1}\Big(\prod_{e\in E}\alpha_{r}\Big)\times
\lim_{w\rightarrow0}w^{-1}Z_{G}\Big({\textstyle\frac{\theta}{2\alpha_{e}}},1,w\Big).
\end{equation}
Recall that the multivariate Bollob\'as-Riordan  polynomial  (see \cite{expansionbollobas}) is a generalization of the multivariate Tutte  polynomial to orientable ribbon graphs defined by the expansion,
\begin{equation}
Z_{G}(\beta,q,w)=\sum_{A\subset E}\Big(\prod_{e\in A}\beta_{e}\Big)q^{k(A)}w^{b(A)},
\end{equation}
with $q(A)$ the number of connected components and $b(A)$ the number of boundaries of the spanning graph $(V,A)$. 

In order to deal with the second Symanzik polynomial in the noncommutative case, we now introduce an extension of $Z_{G}(\beta,q,w)$ for ribbon graphs with flags at $q=1$. In the case of ribbon graphs, the flags are attached to the vertices and the cyclic order of flags and half-edges at each vertex matters. For each cyclically oriented subset $I$ of the set of labels of the flags, we introduce an independent variable $w_{I}$. Cyclically ordered subsets $I$ are defined as sequences of different labels up to a cyclic permutation. Then, each boundary of a graph with the orientation induced by the graph, defines a cyclically ordered subset of the set of labels of the flags, by listing the flags in the order they appear on the boundary. Accordingly, a variable $w_{I}$ is attached to each boundary. 

\begin{definition}
For an orientable ribbon graph $G$ with flags ${\Xi}_{G}(\beta_{e},w_{I})$  is defined by the expansion
\begin{equation}
{\Xi}_{G}(\alpha_{e},\beta_{e},w_{I})=\sum_{A\subset E}\,\Big(\prod_{e\notin E}\alpha_{e}\prod_{e\in E}\beta_{e}\prod_{
\mbox{\tiny boundaries}}w_{I_{n}}\Big),
\end{equation}
where $I_{n}$ are the cyclically ordered sets of flags attached to each of the connected component of the boundary of the spanning graph $(V,A)$. 
\end{definition}

We recover $Z_{G}(\beta_{e},1,w)$ by setting $w_{I}=w$ and $\alpha_{e}=1$, but the information pertaining to $q$ is lost except for planar graphs. Indeed, in this case the genus of any subgraph is still 0 so that $|V|-|A|+b(A)=2k(A)$ and thus $Z_{G}(\beta_{e},q,w)=q^{|V|/2}Z_{G}(q^{-\frac{1}{2}}\beta_{e},q^{\frac{1}{2}}w)$.

The polynomial ${\Xi}_{G}(\alpha_{e},\beta_{e},w_{I})$ obeys the contraction/deletion rules for any semi-regular edges (i.e. all types of edges except self-loops). The structure of the flags of $G-e$ is left unchanged, but less variables $w_{I}$ enter the polynomial since the number of boundaries decreases. For $G/e$, the flags attached to the vertex resulting from the contraction are merged respecting the cyclic order of flags and half-edges attached to the boundary of the subgraph made of the contracted edge only.

\begin{proposition}
The polynomial ${\Xi}_{G}(\alpha_{e},\beta_{e},w_{I})$ obeys the contraction/deletion rule for a semi-regular edge,
\begin{equation}
{\Xi}_{G}(\alpha_{e},\beta_{e},w_{I})=\alpha_{e}\,{\Xi}_{G-e}(\alpha_{e'\neq e},\beta_{e'\neq e},w_{I})+\beta_{e}\,{\Xi}_{G/e}(\alpha_{e'\neq e},\beta_{e'\neq e},w_{I}).
\end{equation}
\end{proposition}

This follows from gathering in the expansion of $\Xi_{G}(\alpha_{e},\beta_{e},w_{I})$ the terms that contain $e$ and those that do not. The contraction/deletion rule may be extended to any edge provided we introduce vertices that are surfaces with boundaries as in \cite{expansionbollobas}.

The second interesting property of ${\Xi}_{G}(\alpha_{e},\beta_{e},w_{I})$ lies in its invariance under duality. Recall that for a connected ribbon graph $G$ with flags, its dual $G^{\ast}$ is defined by taking as vertices the boundaries of $G$, with flags and half-edges attached in the cyclic order following the orientation of the boundary induced by that of $G$.

\begin{proposition}
For a connected graph $G$ with dual $G^{\ast}$,
\begin{equation}
{\Xi}_{G}(\alpha_{e},\beta_{e},w_{I})=
{\Xi}_{G^{\ast}}(\beta_{e},\alpha_{e},w_{I}).
\end{equation}
\end{proposition}

{\it Proof:}
First, recall that there is a natural bijection between the edges of $G$ and those of $G^{\ast}$. Thus, to a subset $A$ of edges of $G$ we associate a subset $A^{\ast}$ of edges of $G^{\ast}$ which is the image under the previous bijection of the complementary $E-A$. Then, the term corresponding to $A$ in ${\Xi}_{G}(\alpha_{e},\beta_{e},w_{I})$ equals that corresponding to $A^{\ast}$ in ${\Xi}_{G^{\ast}}(\beta_{e},\alpha_{e},w_{I})$. The only non trivial part in the last statement is the equality of the boundary terms in $G$ and $G^{\ast}$, which is best understood by embedding $G$ in a surface $\Sigma$. Then, the spanning graph $(V^{\ast},A^{\ast})$, viewed as discs joined by ribbons, is homeomorphic to $\Sigma-(V,A)$, with the orientation reversed. Accordingly, they have the same boundary. 
  
\qed

This relation may also be extended to non connected graphs at the price of introducing again vertices that are surfaces with holes. For example, the dual of a disjoint union of $n$ vertices is the vertex made of a sphere with $n$ holes. For a regular edge, the duality exchanges contraction (resp. deletion) in $G$ with deletion (resp. contraction) in $G^{\ast}$. In the case of the deletion of a bridge in $G$, we have to contract a self-loop in $G^{\ast}$, thus leading to vertices that are surfaces with holes. Note that this implies a duality for the multivariate Bollob\'as-Riordan polynomial only at the special point $q=1$, in agreement with the fact that the duality for the Bollob\'as-Riordan polynomial only holds when its arguments lies on a hypersurface \cite{chmutov}.

Finally, let us come to the relation with the second Symanzik polynomial in NCQFT. For a given connected graph with momenta $p_{i}$ such that $\sum_{i}p_{i}=0$ attached to the flags, we decompose the latter polynomial into real and imaginary part,
\begin{equation}
V^{\star}_{G}(\alpha_{e},\theta,p_{i})={\cal X}^{\star}_{G}(\alpha_{e},\theta,p_{i})+\mathrm{i}\,
{\cal Y}^{\star}_{G}(\alpha_{e},\theta,p_{i}).
\end{equation}

Consider real variables $w_{i}$ and define $w_{I}=\sum_{i}w_{i}$ for any cyclically oriented subset of flags, Then, expand $(\theta/2)^{|E|-|V|}\,{\Xi}_{G}(2\alpha_{e}/\theta,\theta w_{I}/2)$ to the first two orders  at $w_{i}=0$,
\begin{equation}
(\theta/2)^{|E|-|V|}\,{\Xi}_{G}(2\alpha_{e}/\theta,\theta w_{I}/2)=A\Big(\sum_{i}w_{i}\big)+\sum_{i\neq j}B_{ij}w_{i}w_{j}+O\Big(w^{3}\Big).
\end{equation}
The first order term reproduces the first Symanzik polynomial
\begin{equation}
U^{\star}_{G}(\alpha_{e},\theta)=A,
\end{equation}
whereas the second order terms  yields the real part of the second Symanzik polynomial,
\begin{equation}
{\cal X}^{\star}_{G}(\alpha_{e},\theta,p_{i})={\textstyle -\frac{1}{2}}\sum_{i\neq j}A_{ij}\,p_{i}\cdot p_{j}.
\end{equation}
To obtain the imaginary part, consider the variables
\begin{equation}
w_{I}={\textstyle \frac{1}{2}}\sum_{i<j}p_{i}\cdot\Theta p_{j}
\end{equation}
if $I$ contain all the flags and $w_{I}=0$ otherwise. The previous definition involves a choice of a total order on $I$ compatible with its cyclic structure, but momentum conservation $\sum_{i}p_{i}=0$ implies that $w_{I}$ does not depend on this choice. Then,
\begin{equation}
{\cal Y}^{\star}_{G}(\alpha_{e},\theta,p_{i})=(\theta/2)^{|E|-|V|}\,{\Xi}_{G}(2\alpha_{e}/\theta,w_{I}).
\end{equation}
As a consequence of their expressions in terms of ${\Xi}_{G}(\alpha_{e},\beta_{e},w_{I})$,  the noncommutative Symanzik polynomials obey contraction/deletion rules for regular edges and duality relations. For example, the duality for the first Symanzik polynomial reads
\begin{equation}
(\theta/2)^{|V|}\,U^{\star}_{G}(\alpha_{e},\theta)=(\theta/2)^{|V^{\ast}|}\,\Big(\prod_{e\in E}\frac{2\alpha_{e}}{\theta}\Big)\,U^{\star}_{G^{\ast}}\Big({{\theta^{2}}/{\alpha_{e}}},\theta\Big).
\end{equation}
Beware that $G^{\ast}$ is the  dual graph whereas the star on polynomials
such as $U^{\star}$ and $V^{\star}$ refer to the Moyal product.
Analogous relations, though slightly more cumbersome, can be written for the second Symanzik polynomial.

Still an other way to categorify and regularize in the infrared
is to introduce harmonic potentials on the edges rather than the vertices, leading to propagators
based on the Mehler rather than the heat kernel. This is the so-called \emph{vulcanization}. An extensive
study of the corresponding commutative and noncommutative polynomials is 
under way as a companion paper \cite{KRTZ}.

\bigskip

\noindent
{\bf Acknowledgments}

We thank J. Ellis-Monaghan for introducing us to Bollob\'as-Riordan polynomials
and R\u{a}zvan Gur\u{a}u and Fabien Vignes -Tourneret 
for interesting discussions at an early stage of this work.

\end{document}